%% file: sample-acmsmall.tex
\documentclass[acmsmall]{acmart}
\input{macros}

\AtBeginDocument{%
  \providecommand\BibTeX{{%
    \normalfont B\kern-0.5em{\scshape i\kern-0.25em b}\kern-0.8em\TeX}}}

\newcommand{\mybox}[4]{
    \begin{figure}[h]
        \centering
    \begin{tikzpicture}
        \node[anchor=text,text width=\columnwidth-1.2cm, draw, rounded corners, line width=1pt, fill=#3, inner sep=5mm] (big) {\\#4};
        \node[draw, rounded corners, line width=.5pt, fill=#2, anchor=west, xshift=5mm] (small) at (big.north west) {#1};
    \end{tikzpicture}
    \end{figure}
}

\setcopyright{acmcopyright}
\copyrightyear{2018}
\acmYear{2018}
\acmDOI{XXXXXXX.XXXXXXX}

\acmJournal{JACM}
\acmVolume{37}
\acmNumber{4}
\acmArticle{111}
\acmMonth{8}

\begin{document}

\title{A Survey on Automated Software Vulnerability Detection Using Machine Learning and Deep Learning}

\author{Nima Shiri harzevili}
\email{nshiri@yorku.ca}
\affiliation{%
  \institution{York University}
  \streetaddress{4700 Keele St.}
  \city{North York}
  \state{Ontario}
  \country{Canada}
  \postcode{M3J 1P3}
}

\author{Alvine Boaye Belle}
\affiliation{%
  \institution{York University}
  \streetaddress{4700 Keele St.}
  \city{North York}
  \country{Canada}}
\email{alvine.belle@lassonde.yorku.ca}

\author{Junjie Wang}
\affiliation{%
  \institution{Institute of Software, Chinese Academy of Sciences}
  \city{Beijing}
  \country{China}}
  \email{junjie@iscas.ac.cn}

\author{Song Wang}
\affiliation{%
  \institution{York University}
  \streetaddress{4700 Keele St.}
  \city{North York}
  \country{Canada}}
\email{wangsong@yorku.ca}

\author{Zhen Ming (Jack) Jiang}
\affiliation{%
  \institution{York University}
  \streetaddress{4700 Keele St.}
  \city{North York}
  \country{Canada}}
  \email{zmjiang@eecs.yorku.ca}

  \author{Nachiappan Nagappan}
\affiliation{%
  \institution{Meta}
  \city{Seattle}
  \country{USA}}
  \email{nachiappan.nagappan@gmail.com}

\begin{abstract}
Software vulnerability detection is critical in software security because it identifies potential bugs in software systems, enabling immediate remediation and mitigation measures to be implemented before they may be exploited. Automatic vulnerability identification is important because it can evaluate large codebases more efficiently than manual code auditing. Many Machine Learning (ML) and Deep Learning (DL) based models for detecting vulnerabilities in source code have been presented in recent years. However, a survey that summarises, classifies, and analyses the application of ML/DL models for vulnerability detection is missing. It may be difficult to discover gaps in existing research and potential for future improvement without a comprehensive survey. This could result in essential areas of research being overlooked or under-represented, leading to a skewed understanding of the state of the art in vulnerability detection. This work address that gap by presenting a systematic survey to characterize various features of ML/DL-based source code level software vulnerability detection approaches via five primary research questions (RQs). Specifically, our RQ1 examines the trend of publications that leverage ML/DL for vulnerability detection, including the evolution of research and the distribution of publication venues. RQ2 describes vulnerability datasets used by existing ML/DL-based models, including their sources, types, and representations, as well as analyses of the embedding techniques used by these approaches. RQ3 explores the model architectures and design assumptions of ML/DL-based vulnerability detection approaches. RQ4 summarises the type and frequency of vulnerabilities that are covered by existing studies. Lastly, RQ5 presents a list of current challenges to be researched and an outline of a potential research roadmap that highlights crucial opportunities for future work.


\end{abstract}

                    

\maketitle

\input{Sections/introduction}
\input{Sections/related}

\input{Sections/methodology}

\input{Sections/RQ1}

\input{Sections/RQ2}
\input{Sections/RQ3}
\input{Sections/RQ4}

\input{Sections/RQ6}

\input{Sections/Limitations}

\input{Sections/conclusion}


\bibliographystyle{ACM-Reference-Format}
\bibliography{sample-acmsmall}

\appendix

\end{document}

%% file: macros.tex
\usepackage{caption}
\usepackage{subcaption}
\usepackage[utf8]{inputenc} 
\usepackage{geometry} 
\usepackage{tikz}
\usepackage{lipsum}
\usepackage{soul}
\usepackage{balance}
\usepackage{hyperref}
\usepackage{graphicx}
\usepackage{soul}
\usepackage{hyperref}
\usepackage{multirow}
\usepackage{fancybox}
\usepackage{enumitem}
\usepackage{graphicx}
\usepackage{color}
\usepackage{float}
\usepackage{xcolor}
\usepackage{array}
\usepackage{amsmath}
\usepackage{pgfplots}
\usepackage{pgfplotstable}
\usepackage{xspace}
\usepackage{url}
\usepackage{pgfkeys}
\usepackage{tikz}
\usepackage{caption}
\usepackage{pgfplots}
\usepackage{listings}
\usepackage{color}
\usepackage{graphicx}
\definecolor{dkgreen}{rgb}{0,0.6,0}
\definecolor{gray}{rgb}{0.5,0.5,0.5}
\definecolor{mauve}{rgb}{0.58,0,0.82}
\pgfplotsset{compat=1.16}
\usepackage{pgf-pie}
\usetikzlibrary{pgfplots.statistics,calc}
\usepackage{algorithm}
\usepackage{algorithmic}
\usepackage{subcaption}
\usepackage{listings}
\usetikzlibrary{shapes.geometric, arrows}

\definecolor{songcolor}{RGB}{191,191,191}



%% file: Sections/introduction.tex
\section{Introduction}
\label{sec:intro}
Automatic detection of software security vulnerabilities is a critical component of assuring software security. Machine Learning (ML) and Deep Learning (DL) breakthroughs have sparked great interest in employing these models to discover software vulnerabilities in general software systems..~\cite{wang2018deep,cheng2021deepwukong,li2021vulnerability,le2021deepcva,yan2021han}. {ML/DL models excel at discovering subtle patterns and correlations from large datasets. They can automatically extract meaningful features from raw data, such as source code, and identify hidden patterns that may indicate software vulnerabilities. This capability is crucial in vulnerability detection, as vulnerabilities often involve subtle code characteristics and dependencies. Also, ML/DL models can handle a wide range of data types and formats, including source code~\cite{dam2018automatic,
dam2017automatic,
shippey2019automatically,
wang2016automatically,
jeon2021autovas,
tian2020bvdetector,
liu2020cd,
yamaguchi2013chucky}, textual information~\cite{hoang2019deepjit}, and numerical features such as commit characteristics~\cite{pascarella2019fine,
Vuldiggerpaper}. They can process and analyze these data representations to detect vulnerabilities effectively. This flexibility allows researchers to leverage various sources of data and incorporate different features for comprehensive vulnerability detection.} The overall process to leverage ML/DL models for software vulnerability detection is as follows: \noindent \textbf{Data collection:} The first step toward building a vulnerability detection model is to collect relevant vulnerable data for training the models. There are multiple sources for vulnerability detection datasets (we elaborate on this in RQ2), the researchers either use benchmark data~\cite{le2018maximal,
zou2019mu,
cao2022mvd,
ghaffarian2021neural,
wu2021peculiar,
ziems2021security,
zhuang2020smart,
li2021sysevr,
filus2020random,
li2018vuldeepecker,
Vuldiggerpaper} or collect from the open source~\cite{fu2022linevul,
chen2021neural,
perl2015vccfinder,
riom2021revisiting,
ni2022best} based on the requirements and the type of vulnerabilities. \noindent \textbf{Data representation:} Once the data is collected, it needs to be preprocessed to prepare it for training. The preprocess includes using appropriate representation techniques, i.e., graph/tree representation~\cite{zou2019mu,
cao2022mvd,
ghaffarian2021neural,
wu2021peculiar,
lin2017poster,
zhuang2020smart,
li2017software,
lin2019software,
li2021sysevr,
duan2019vulsniper,
zheng2021vulspg,
zhuang2022just,
cheng2022path,
liu2021smart}, token representation~\cite{3429444,
8594942,
3327890,
hin2022linevd,
le2018maximal,
chen2021neural,
scandariato2014predicting,
ziems2021security,
filus2020random}, or using commit characteristics. \noindent \textbf{Embedding:} This step involves converting the source code representation into numerical format~\cite{3429444,
8594942,
3327890,
hin2022linevd,
fu2022linevul,
le2018maximal,
cao2022mvd,
ghaffarian2021neural,
lin2017poster,
scandariato2014predicting,
perl2015vccfinder}(vectors or embeddings) that can be utilized by machine learning or deep learning models for vulnerability detection. \noindent \textbf{Model selection and architecture design:} An suitable ML/DL model must be chosen based on the software vulnerability detection task. This can include everything from simple ML algorithms like SVM or Random Forests~\cite{chen2020machine,
sabetta2018practical,
zhou2017automated} to more advanced DL architectures like CNNs~\cite{hoang2019deepjit,
yan2021han,
8594942} or RNNs. The architecture of the model is intended to extract significant characteristics and patterns from the input data. \noindent \textbf{Training:} In the training phase, the vulnerability detection dataset is separated into training and validation sets, and the model learns from labeled data. The model's parameters are updated iteratively depending on the prediction errors, using optimization techniques such as gradient descent. \noindent \textbf{Evaluation and validation:} Once the training is finished, the model's performance is evaluated using a separate test dataset. Various metrics such as accuracy, precision, recall, and F1 score are calculated to assess the model's effectiveness in detecting vulnerabilities. The model may also be validated against real-world vulnerabilities to measure its practical utility.

Although many studies have utilized ML/DL to detect software vulnerabilities, there has not been a comprehensive review to consolidate the various approaches and characteristics of these techniques. Conducting such a systematic survey would be beneficial for practitioners and researchers to gain a better understanding of the current state-of-the-art tools for vulnerability detection, and could serve as inspiration for future studies. 
This study conducts a detailed and comprehensive survey to review, analyze, describe, and classify vulnerability detection papers from different perspectives. We analyzed 67 articles published in 37 flagship SE journals and conferences from 2011 to 2022. 
We investigated the following research questions (RQs) in this study:
\begin{itemize}
    \item \textbf{RQ1: What is the trend of studies using ML/DL models for vulnerability detection?}
    \begin{itemize}
        \item RQ1.1. What are the trends of studies in software vulnerability detection studies over time?
        \item RQ1.2. What is the distribution of the publication venues?
    \end{itemize}
    \item \textbf{RQ2: What are the characteristics of experiment datasets used in software vulnerability detection?}
    \begin{itemize}
    \item RQ2.1. What is the source of data?
    \item RQ2.2. What are the types of data used in primary studies?
    \item RQ2.3. How input data are represented?
    \item RQ2.4. How input data are embedded for feature space?
    \end{itemize}
    \item \textbf{RQ3. What are the different ML/DL models used for vulnerability detection?}
    \item \textbf{RQ4. What are the most frequent type of vulnerabilities covered in these studies?}
    \item \textbf{RQ5. What are possible challenges and open directions in software vulnerability detection?}
\end{itemize}

This paper makes the following contributions:

\begin{itemize}
    \item We thoroughly analyzed 67 relevant studies that used ML/DL techniques to detect security vulnerabilities regarding publication trends, distribution of publication venues, and types of contributions.
    \item We conducted a comprehensive analysis to understand the dataset, the processing of data, data representation, model architecture, model interpretability, and the types of involved vulnerabilities of these ML/DL-based vulnerability detection techniques.
    \item We provided a classification of ML/DL models used in vulnerability detection based on their architectures and analysis of technique selection strategy on these models.
    \item We discuss distinct technical challenges of using ML/DL techniques in vulnerability detection and outline key future directions. 
    \item We have shared our results and analysis data as a replication package\footnote{https://colab.research.google.com/drive/1O42duwz34H3fRoyfA37EU6Ig2u16R1Lb?usp=sharing} to allow other researchers easily follow this paper and extend it. 
\end{itemize}

{We believe that this work is useful for researchers and practitioners in the field of software engineering and cybersecurity, particularly those with an interest in software vulnerability detection and mitigation. In addition, the findings of our systematic survey may also be useful to policymakers, software vendors, and other stakeholders who are concerned with improving software security and reducing the risk of cyberattacks. These individuals may use the insights provided by the review to inform their decisions about software development, procurement, and risk management.}

The remaining part of this paper is organized as follows: Section~\ref{related} summarizes existing studies focusing on proposing a systematic survey for software vulnerability detection.
Section 2 presents related work on systematic surveys for software vulnerability detection using ML/DL techniques.
Section 3 presents the research methodology proposed in this paper for paper collection and criteria for including and excluding studies. Section 4 addresses research questions and corresponding results. Section 5 discusses the possible limitations of this systematic survey. Finally, section 6 discusses the conclusion and future directions. 

%% file: Sections/related.tex
\section{Background and Related Work}
\label{related}

In this section of the paper, we first provide a background on the definition of vulnerability and the different steps in software vulnerability detection. Then we discuss the related surveys and highlight their differences compared to our survey.

\subsection{Background}

Software vulnerability management is now essential for guaranteeing the security and integrity of software systems~\cite{chang2011trend, foreman2019vulnerability, raza2022threat, walkowski2021vulnerability}. Given the increasing reliance on software for many critical processes such as financial transactions, the frequency of vulnerabilities poses serious risks..~\cite{lu2021neucheck, he2020smart}, autonomous driving~\cite{gao2021autonomous, luo2019localization}, and mission-critical systems~\cite{goseva2017experience, harzevili2022characterizing}. Software vulnerabilities can be exploited by malicious entities to gain unauthorized access, compromise sensitive information, or disrupt services if they go undetected or ignored.~\cite{harzevili2022characterizing}. As a result, excellent software vulnerability management is crucial to handling these risks, preserving user privacy~\cite{anthonysamy2017privacy}, maintaining system availability, and assuring software application trustworthiness~\cite{medeiros2023trustworthiness}. By proactively detecting, analyzing, and remediating vulnerabilities, organizations may strengthen their software systems against changing cybersecurity threats~\cite{aslan2023comprehensive} and adhere to industry best practices for safe software development and deployment.

There are multiple steps in software vulnerability management including vulnerability detection~\cite{cao2022mvd}, vulnerability analysis~\cite{kudjo2020effect}, and vulnerability remediation~\cite{le2021survey}. In the following subsections, we elaborate on each step in detail.

\subsubsection{Vulnerability detection}
Vulnerability detection is critical in the overall process of managing software vulnerabilities~\cite{cao2022mvd, jeon2021autovas,
 tian2020bvdetector,
 liu2020cd,
 wang2020combining,
 cheng2021deepwukong,
 zou2019mu,
 cao2022mvd,
 ziems2021security,
 lin2019software,
 li2021sysevr}. It comprises detecting and investigating possible security weaknesses in software systems that attackers may exploit. There are several traditional techniques commonly used for vulnerability detection:

\noindent \textbf{Manual Code Auditing} \cite{staron2020using, bacchelli2013expectations, carlsson2005software, shar2012auditing, shar2010auditing}: In this method, human experts examine the source thoroughly with the goal of manually detecting coding flaws, unsafe procedures, and possible vulnerabilities. Manual code review is time-consuming and requires the knowledge of qualified developers or security analysts. However, it provides for a thorough grasp of the code and can reveal subtle bugs that automated tools may overlook.

\noindent \textbf{Static Analysis}~\cite{infer, flawfinder, rats, cppcheck, SpotBugs, lattner2008llvm}: Static analysis involves using automated tools to analyze the source code or compiled binaries without executing the software. It examines the code structure, identifies potential coding issues, and detects common vulnerabilities such as buffer overflows~\cite{harzevili2022characterizing}, injection attacks, and insecure data handling. Static analysis tools employ various techniques like data flow analysis, control flow analysis, and pattern matching to identify potential vulnerabilities. They can help scale vulnerability detection efforts by analyzing large codebases efficiently.

\noindent \textbf{Dynamic Analysis}~\cite{nethercote2007valgrind, lehmann2019wasabi, cadar2008klee}: The goal of dynamic analysis is to evaluate the behavior of software while it is running. Running the software in a controlled environment or through automated tests while monitoring its execution and interactions with system resources is what it entails. Dynamic analysis can detect bugs in input validation~\cite{ kim2019rvfuzzer}, access control, and error handling. This approach can identify vulnerabilities that static analysis alone cannot detect by analyzing the real-time behavior of the software. However, the dynamic analysis may have constraints in terms of significant system overhead~\cite{yong2005using}.

\subsubsection{Vulnerability analysis}
After the detection of vulnerabilities, the subsequent step in software vulnerability management is vulnerability analysis and assessment~\cite{le2019automated, suciu2022expected, jacobs2023enhancing, yin2022real, le2022survey, mahor2022mobile, frei2006large}. This step involves further examining the identified vulnerabilities to assess their severity, impact, and potential exploitability.

\noindent \textbf{Severity}: Accurately assessing software vulnerabilities is vital for several reasons. Firstly, it allows organizations to prioritize their response based on the severity of the vulnerabilities. Severity refers to the potential impact a vulnerability could have if exploited~\cite{kudjo2019improving, chen2020automatic, kudjo2020effect, tan2020bug}. By accurately assessing the severity, organizations can focus their attention on high-severity vulnerabilities that pose significant risks to the security and functionality of the software system.

\noindent \textbf{Impact}: Secondly, accurately assessing vulnerabilities helps determine the potential impact they may have on the organization~\cite{gawron2018automatic, gong2019joint, chen2010categorization, jiang2020approach}. The term \textit{impact} refers to the repercussions of exploiting a vulnerability, such as denial of service~\cite{harzevili2022characterizing} or data breaches~\cite{aaltonen2021does}. By understanding the potential impact, organizations can make informed decisions regarding the urgency and priority of remediation efforts.

\noindent \textbf{exploitability}: Furthermore, accurately assessing vulnerabilities aids in understanding their potential exploitability~\cite{chen2019using, chen2019vest, bozorgi2010beyond}. This entails determining the possibility that an attacker will be successful in exploiting the vulnerability to infiltrate the software system. Organizations can estimate the amount of risk associated with each vulnerability and invest resources accordingly by evaluating criteria such as the ease of exploitation and the availability of exploit techniques.

\subsubsection{Vulnerability remediation}
The process of addressing detected software vulnerabilities by different techniques such as patching, code modification, and repairing is referred to as software vulnerability remediation~\cite{canfora2022patchworking, piantadosi2019fixing, chen2022neural, bhandari2021cvefixes}. The fundamental goal of remediation is to eliminate or mitigate vulnerabilities in order to improve the software system's security and dependability. One common approach to vulnerability remediation is applying patches provided by software vendors or open-source communities~\cite{xia2023automated, li2022generating, gissurarson2022propr}. Patches are updates or fixes that address specific vulnerabilities or weaknesses identified in a software system. 

\subsubsection{ML/DL for software vulnerability detection}

By utilizing data analysis, pattern recognition, and machine-driven learning for finding software security vulnerabilities, ML/DL approaches have revolutionized software vulnerability detection.~\cite{wang2018deep,cheng2021deepwukong,li2021vulnerability,le2021deepcva,yan2021han}. These techniques improve the accuracy and efficiency of vulnerability detection, potentially allowing for automated detection, faster analysis, and the identification of previously undisclosed vulnerabilities.

One common application of ML/DL in vulnerability detection is the classification of code snippets~\cite{dam2017automatic, shippey2019automatically, wang2016automatically, jeon2021autovas}, software binaries~\cite{phan2017convolutional,
nguyen2020deep,
yan2021han,
huang2021hunting}, or code changes mined from open-source repositories such as GitHub or CVE~\cite{sabetta2018practical,
wang2020combining,
liu2019deepbalance,
pradel2018deepbugs,
hoang2019deepjit,
zhou2019devign,
dinella2020hoppity,
3360588}. ML models can be trained on labeled datasets, where each sample represents a known vulnerability or non-vulnerability. These models then learn to generalize from the provided examples and classify new instances based on the patterns they have learned. This method allows for automatic vulnerability discovery without the need for manual examination, considerably lowering the time and effort necessary for analysis. 

ML/DL models for detecting software vulnerabilities have promising advantages over traditional methodologies. Each benefit is discussed in depth in the next paragraph. 

\noindent \textbf{Automation}: Automation is a significant advantage. ML models can automatically scan and analyze large codebases, network traffic logs, or system configurations, flagging potential vulnerabilities without requiring human intervention for each individual case~\cite{chakraborty2021deep}. This automation speeds up the detection process, allowing security teams to focus on verifying and mitigating vulnerabilities rather than manual analysis.

\noindent \textbf{Performance}: ML/DL approaches offer faster analysis. Traditional vulnerability detection methods rely on manual inspection or the application of predefined rules~\cite{staron2020using, bacchelli2013expectations, carlsson2005software, shar2012auditing, shar2010auditing}. In contrast, ML/DL approaches can evaluate enormous volumes of data in parallel and generate predictions fast, dramatically shortening the time necessary to find vulnerabilities.

\noindent \textbf{Detection effectiveness}: ML/DL models can uncover previously unknown vulnerabilities, commonly known as zero-day vulnerabilities~\cite{bilge2012before}. These models may uncover signs of vulnerabilities even when they have not been specifically trained on them by learning patterns and generalizing from labeled data. This capability improves the overall security posture by helping to identify and address unknown weaknesses in software before they are exploited by attackers~\cite{abri2019can}.

\subsection{Related work}
There have been several existing survey papers on software vulnerabilities in the literature. In this section, we analyze the existing papers based on different aspects as shown in Table~\ref{relatedTable}.

\input{Tables/table1}

The columns in the table represent different aspects of the surveys, such as the data source used, representation, feature embedding, ML/DL models, vulnerability types, and interpretability of ML/DL models. \textit{Data Source} indicates whether the survey reviewed vulnerability detection data sources. \textit{Representation} discusses whether the survey considered source code representation in its analysis. \textit{Embedding} deals with whether the survey is analyzed feature embedding in its analysis. The table also considers ML/DL models in the sixth column as \textit{ML Models}. The table also checks whether the survey considers vulnerability types based on Common Weakness Enumeration (CWE) number. The last column indicates whether the survey takes into account the interpretability of ML/DL models.



Ghaffarian et al. \cite{ghaffarian2017software} is the closest survey to ours when it comes to data-driven security vulnerability detection. 
{In their survey, they analyzed data-driven software vulnerability detection from various aspects including \textit{Data Sources}, \textit{Representation}, \textit{Embedding} types, and different ML/DL models as shown in Table~\ref{relatedTable}}. However, there are a couple of differences compared to our work. 
Specifically, this work also surveys vulnerability detection from the following aspects: \textbf{Comprehensive Coverage}: Understanding the many sorts of vulnerabilities allows researchers to create and develop effective vulnerability detection models that can thoroughly discover security vulnerabilities. To guarantee that their detection systems cover as many vulnerability types as possible, researchers must be familiar with the various methods of attack and potential weaknesses in software systems. \textbf{Customization of Detection Techniques}: Different sorts of vulnerabilities necessitate distinct detection methods. To build specialized detection systems that can discover certain types of vulnerabilities, researchers must first understand the subtleties of each vulnerability type. \textbf{Prioritization of Mitigation Efforts}: Researchers can prioritize mitigation efforts depending on the severity and effect of each vulnerability by understanding the many types of vulnerabilities. Critical vulnerabilities that pose the greatest danger to the system or organization can be prioritized by researchers. \textbf{Better Understanding of Attack Patterns}: Understanding the different types of vulnerabilities provides researchers with insights into the different attack patterns used by attackers. This knowledge helps researchers design detection techniques that can detect not only known attack patterns but also new, unknown patterns. 
\textbf{Interpretability} refers to the ability to explain how a model makes a particular decision or prediction. This is particularly important in the context of software vulnerability detection because security researchers need to be able to understand why a model is flagging a particular piece of code as potentially vulnerable. 
Additionally, interpretability can help improve trust in the model's predictions. If developers and security researchers can understand how a model is making its decisions, they are more likely to trust its output and take appropriate actions based on its recommendations. 


Triet et al.~\cite{le2021survey} reviewed data-driven vulnerability assessment and prioritization studies. They conduct a review of prior research on software assessment and prioritization that leverages machine learning and data mining methods. They examine various types of research in this area, discuss the strengths and weaknesses of each approach, and highlight some unresolved issues and potential areas for future research. The major difference to ours is that we review vulnerability detection approaches while they survey assessment and prioritization techniques. Vulnerability detection, vulnerability assessment, and vulnerability prioritization are all important components of the vulnerability management life-cycle, but they involve different stages of the vulnerability management process. Our work focuses on Vulnerability detection which refers to the process of identifying potential vulnerabilities in software systems. The goal of vulnerability detection is to identify all vulnerabilities that exist within the system, regardless of their severity. Vulnerability assessment, on the other hand, involves evaluating the severity and potential impact of each identified vulnerability. This assessment can involve analyzing factors such as the likelihood of the vulnerability being exploited and the potential harm that could result. Vulnerability prioritization involves ranking the identified vulnerabilities based on their level of risk or criticality. This ranking is typically based on the results of the vulnerability assessment, as well as other factors such as the availability of resources to address the vulnerabilities. 

Lin et al. \cite{lin2020software} examined the literature on using deep learning and neural network-based approaches for detecting software vulnerabilities. There are a couple of differences compared to our survey. First, the study of conventional source code representation techniques (Static code attributes) for software vulnerability detection. In our survey, we neglect to review such representation techniques for a couple of reasons. Static code attributes, such as code length or complexity, may not be effective for vulnerability detection because they do not capture the dynamic behavior of the code at runtime. Vulnerabilities can manifest themselves in unexpected ways that are not apparent in the static code, making it difficult to detect them through static analysis alone. Additionally, static code attributes may not be able to capture the context of the code, which is important for understanding how the code interacts with other components in a system. Finally, static analysis tools may produce a high rate of false positives, which can be time-consuming to verify and may cause developers to ignore important vulnerabilities. Second, we examine the trend analysis of papers published in software vulnerability detection in a journal and conference papers because it provides a comprehensive understanding of the publishing patterns in a particular field or area of research. The trend analysis can shed light on the distribution of research output across various publication venues and the shifting preferences of researchers and authors. This information can be useful for stakeholders such as publishers, academic institutions, and researchers in making strategic decisions related to publishing, funding, and research collaborations.

Zeng et al. \cite{zeng2020software} discussed the increasing attention towards exploitable vulnerabilities in software and the development of vulnerability detection methods, specifically the application of ML techniques. The paper reviews 22 recent studies that use deep learning to detect vulnerabilities and identifies four \textit{game changers} that have significantly impacted this field. The survey further compares the game changers based on different aspects in software vulnerability detection including data source, feature representation, DL models, and detection granularity. There are a couple of differences compared to our survey. First, we analyze the trend patterns of papers on software vulnerability detection that have been published in journals and conferences. This analysis helps us gain a thorough comprehension of the publication trends in a specific area of research or field.
Second, we cover more aspects of software vulnerability detection. While they only cover data source, feature representation, DL models, and detection granularity, we cover more aspects including vulnerability types and interpretability of ML/DL models. Additionally, we provide a more granular analysis of different aspects.

Kritikos et al.~\cite{kritikos2019survey} and Sun et al.~\cite{sun2018data} focused on cybersecurity and aim to improve cyber resilience. Sun et al.~\cite{sun2018data} discussed the paradigm change in understanding and protecting against cyber threats from reactive detection to proactive prediction, with an emphasis on new research on cybersecurity incident prediction systems that use many types of data sources. Kritikos et al.~\cite{kritikos2019survey} discusses the challenges of migrating applications to the cloud and ensuring their security, with a focus on vulnerability management during the application lifecycle and the use of open-source tools and databases to better secure applications. While the topics of the two abstracts are different, they share a common goal of improving cybersecurity and resilience. Both highlight the importance of proactive measures to prevent or mitigate cyber threats, rather than relying solely on reactive detection and response. Additionally, both highlight the importance of utilizing various data sources and tools to improve cybersecurity measures. While both approaches aim to improve the security of applications, they differ in their focus and techniques used. They mainly focus on providing guidance and tools to support vulnerability management during the application lifecycle, while in our survey, we focus on software vulnerability detection using ML/DL techniques on source code which aim at automating the identification of vulnerabilities in the source code.

Khan et al.~\cite{khan2018review} focused on Vulnerability Assessment, which is the process of finding and fixing vulnerabilities in a computer system before they can be exploited by hackers. This highlights the necessity for more studies into automated vulnerability mitigation strategies that can effectively secure software systems. On the other hand, vulnerability identification with ML/DL approaches on source code entails analyzing an application's source code in order to spot security flaws. Instead of evaluating the safety of the entire system, this method concentrates on finding vulnerabilities in the code itself.

Nong et al.\cite{nong2022open} explored the open-science aspects of studies on software vulnerability detection and argued there is a dearth of research on problems of open science in software engineering, particularly with regard to software vulnerability detection. The authors conducted an exhaustive literature study and identify 55 relevant works that propose deep learning-based vulnerability detection approaches. They investigated open science aspects including availability, executability, reproducibility, and replicability. The study reveals that 25.5\% of the examined approaches provide open-source tools. Furthermore, some open-source tools lack adequate documentation and thorough implementation, rendering them inoperable or unreplicable. The use of unbalanced or intentionally produced datasets causes the approaches' performance to be overstated, rendering them unreplicable.

Chakraborty et al.~\cite{chakraborty2021deep} investigated the performance of cutting-edge DL-based vulnerability prediction approaches in real-world vulnerability prediction scenarios. They find that the performance of the state-of-the-art DL-based techniques drops by more than 50 percent in real-world scenarios. They also discover problems with training data (for example, data duplication and an unrealistic distribution of vulnerable classes) and model selection (for example, simplistic token-based models). Existing DL-based approaches often learn unrelated artifacts instead of features related to the cause of vulnerabilities. The significant difference compared to our survey study is that in our work, we focus on the usage of ML/DL models for software vulnerability detection and characterize the different stages in the pipeline of vulnerability detection. On the other hand, they focus on the issues related to the usage of state-of-the-art DL models for software vulnerability detection. 

Liu et al.~\cite{liu2021reproducibility} discussed the increasing popularity of DL techniques in software engineering research due to their ability to address SE challenges without extensive manual feature engineering. The authors highlight two important factors often overlooked in DL studies: reproducibility and replicability. Reproducibility refers to whether other researchers can obtain the same results using the authors' artifacts, while replicability refers to obtaining similar results with re-implemented artifacts and a different experimental setup. The major difference compared to our study is that we focus on the usage of ML/DL techniques in software vulnerability detection pipelines, while they emphasize replicability and reproducibility of the results reported in software engineering research studies. 

Bi et al.~\cite{bi2023benchmarking} emphasizes the importance of software vulnerability detection techniques as well as the absence of a systematic methodology to evaluate these approaches. The research is the first to look into and describe the current state of software vulnerability detection benchmarking. The assessment examines current literature on vulnerability detection benchmarking, including methodologies employed in technique-proposing publications and empirical research. The survey examines the difficulties associated with benchmarking software vulnerability detection approaches and suggests alternative solutions to these difficulties. They do not, however, give a characterization of datasets, representations, embedding techniques, and models employed in software vulnerability identification, unlike our work.

%% file: Tables/table1.tex
\begin{table}[t!]
\caption{Comparison of contributions between our survey and the existing related surveys/reviews.}
\label{relatedTable}
\renewcommand{\arraystretch}{0.1}
\resizebox{\columnwidth}{!}{
\begin{tabular}{llccccccc}
\toprule
No &  Studies &  Data Source & Representation & Embedding & Models & Vulnerability Types & Interpretability \\
\midrule
1  & Triet et al. \cite{le2021survey}   & $$\checkmark$$ & $\times$     & $\checkmark$  & $\checkmark$ & $\times$     &  $\times$ \\
2  & Ghaffarian et al. \cite{ghaffarian2017software} & $\checkmark$ & $\checkmark$ & $\checkmark$  & $\checkmark$ & $\times$     & $\times$\\
3  & Lin et al. \cite{lin2020software} & $\checkmark$ & $\checkmark$ & $\checkmark$  & $\checkmark$ & $\times$     & $\times$\\
4  & Zeng et al. \cite{zeng2020software}  & $\checkmark$ & $\checkmark$ & $\checkmark$  & $\checkmark$ & $\times$     &  $\times$\\
5  & Semasaba et al. \cite{semasaba2020literature} & $\checkmark$ & $\checkmark$ & $\times$      & $\checkmark$ & $\checkmark$ &  $\times$\\
6  & Sun et al. \cite{sun2018data}  & $\checkmark$ & $\times$     & $\times$      & $\times$     & $\checkmark$ & $\times$\\
7  & Kritikos et al. \cite{kritikos2019survey} & $\checkmark$ & $\times$     & $\times$      & $\times$     & $\checkmark$ &  $\times$\\
8  & Khan et al. \cite{khan2018review}  & $\times$  &  $\times$  &  $\times$   &  $\times$    &  $\times$   & $\times$ \\
9  & Nong et al. \cite{nong2022open}  & $\times$  &  $\times$  &  $\times$   &  $\times$    &  $\times$   & $\times$ \\
10  & Chakraborty et al. \cite{chakraborty2021deep}  & $\times$  &  $\times$  &  $\times$   &  $\times$    &  $\times$   & $\times$ \\
11  & Liu et al. \cite{liu2021reproducibility}  & $\times$  &  $\times$  &  $\times$   &  $\times$    &  $\times$   & $\times$ \\
12  & Bi et al. \cite{bi2023benchmarking}  & $\times$  &  $\times$  &  $\times$   &  $\times$    &  $\times$   & $\times$ \\
12  & Our survey & $\checkmark$ & $\checkmark$ & $\checkmark$  & $\checkmark$ & $\checkmark$ & $\checkmark$ \\
\bottomrule
\end{tabular}}
\end{table}

%% file: Sections/methodology.tex
\section{Methodology of Systematic Survey}
\subsection{Sources of Information}
In this paper, we conducted an empirical study following \cite{keele2007guidelines,petersen2015guidelines}. The purpose of this study is to collect and examine papers from the year 2011 to 2022 focusing on vulnerability detection across various programming languages and source codes using machine learning and deep learning techniques. The period between 2011 and 2022 is an appropriate time interval for reviewing data-driven vulnerability detection for several reasons: 
a) {\textbf{Increase in the volume and diversity of software vulnerabilities}: Over the past decade, there has been a significant increase in the number and diversity of software vulnerabilities that have been discovered and reported~\footnote{https://nvd.nist.gov/general/news}. As of 2021, there exist 150,000 CVE records in the National Vulnerability Database (NVD)\footnote{https://nvd.nist.gov/general/brief-history}. This increase has created a need for more sophisticated and effective methods for vulnerability detection, which has led to the development of new data-driven techniques.
b) \textbf{Advancements in ML/DL and data analytics}: The past decade has seen significant advancements in machine learning, including the development of deep learning algorithms~\cite{goodfellow2020generative, hinton2006fast}, natural language processing techniques~\cite{devlin2018bert, liu2019roberta}, and other data-driven approaches that are highly effective in detecting software vulnerabilities.

While we start collecting papers, we search for relevant research papers from four available databases, which are ScienceDirect, IEEE Xplore, ACM digital library, and Google Scholar. 
\subsection{Search Terms}
From earlier work, we identify key phrases used in the search~\cite{le2021survey,lin2020software,zeng2020software,semasaba2020literature} and our experience with the subject area. The following are the search terms:

\noindent\fbox{%
\parbox{\textwidth}{\textit{vulnerability detection} \textcolor{red}{OR} \textit{security vulnerability detection} \textcolor{red}{OR}
\textit{vulnerability detection using machine learning} \textcolor{red}{OR} \textit{vulnerability detection using deep learning} \textcolor{red}{OR} 
\textit{source code security bug prediction} \textcolor{red}{OR} \textit{source code vulnerability detection} \textcolor{red}{OR} \textit{source code bug prediction}}}

\subsection{Study Selection}
The process of selecting studies to be included in our survey involves the following stages: (1) initially choosing studies based on their title, (2) selecting studies after reviewing their abstracts, and (3) making further selections after reading the full papers. Note that, the initial search results contain entries that are not related to security vulnerability detection. This might be caused by accidental keyword matching. We manually check each paper and remove these irrelevant papers to ensure the quality of our survey dataset. We also observe that there exist duplicate papers among search results  since the same study could be indexed by multiple databases. We then discarded duplicate studies manually. 
To assist the selection of papers that have presented new ML or DL-based models for software vulnerability identification, we provide the following inclusion and exclusion criteria:
\begin{itemize}
\item The studies should have been peer-reviewed 
    \item The studies should have experimental results
    \item The studies should employ an ML or DL technique
    \item The studies improve existing data-driven vulnerability detection techniques
    \item The input to ML/DL models should be either source code, commit, or byte-codes
\end{itemize}

Also, we have the following exclusion criteria to filter out irrelevant papers:

\begin{itemize}
    \item Studies focusing on other engineering domains
    \item Studies addressing static analysis, dynamic analysis, mutation testing, fault localization
    \item Review papers
    \item Studies focusing on vulnerability detection of web and Android applications
    \item Studies belonging to one of the following categories: books, chapters, tutorials, technical reports
    \item Studies using code similarity or clone detection tools
    \item Studies focusing on malware detection on mobile devices, intrusion detection, and bug detection using static code attributes
\end{itemize}

Using these criteria, we narrow down our findings by examining each paper's title, abstract, and contents to get the most relevant and high-quality research papers. To make human effort manageable, we developed a script to automatically get high-quality records close to the software vulnerability detection problem. To summarize, in the first stage, we began with a total of 3,154 papers obtained from the database search. From this initial pool, 880 papers were chosen for further evaluation in the second stage. During the second stage, these papers were reviewed based on their abstracts, resulting in the selection of 116 papers with relevant abstracts. Finally, in the third stage, after reading the full papers, 67 papers were ultimately chosen for inclusion in the study. 

\subsection{Study Quality Assessment}
\label{sec:squality}
For each of the final selected studies, we answered the questions below to assess its quality:
\begin{itemize}
\item Is there a clearly stated research goal related to software vulnerability detection? 
\item Is the proposed vulnerability detection approach used ML or DL techniques?
\item Is there a defined and repeatable technique?
\item Is there any explicit contribution to vulnerability detection?
\item Is there a clear methodology for validating the technique?
\item Are the subject projects selected for validation suitable for the research goals?
\item Are there control techniques or baselines to demonstrate the effectiveness of the vulnerability detection technique?
\item Are the evaluation metrics relevant (e.g., evaluate the effectiveness of the proposed technique) to the research objectives? 
\item Do the results presented in the study align with the research objectives and are they presented in a clear and relevant manner?
 \end{itemize}   


\subsection{Selection Verification}
\label{sec:sverification}
The process of creating a taxonomy for the selected 67 primary studies involves several steps. Initially, the lead author establishes a preliminary taxonomy that groups the studies together based on their research questions. This taxonomy provides a basic framework for organizing the studies in a meaningful and systematic manner. Next, the lead author expands the taxonomy by assigning new papers to the preliminary taxonomy. If a new paper cannot fit into any of the existing categories within the taxonomy, a new category is created that reflects the unique characteristics of that paper. To ensure the accuracy of the taxonomy, the second and third authors (who are not involved in the taxonomy creation process) randomly select 20 papers from the workflow and check the created taxonomies for any discrepancies. They then mark any disagreements they find, and all three authors discuss and resolve these disagreements. Initially, the disagreement rate was 30\%, but after a second round of review and cross-checking of the papers, the authors were able to eliminate all disagreements.

\input{Figures/sources/fig1}

%% file: Figures/sources/fig1.tex
\begin{figure}[t!]
    \centering
     {\includegraphics[width=1\textwidth]{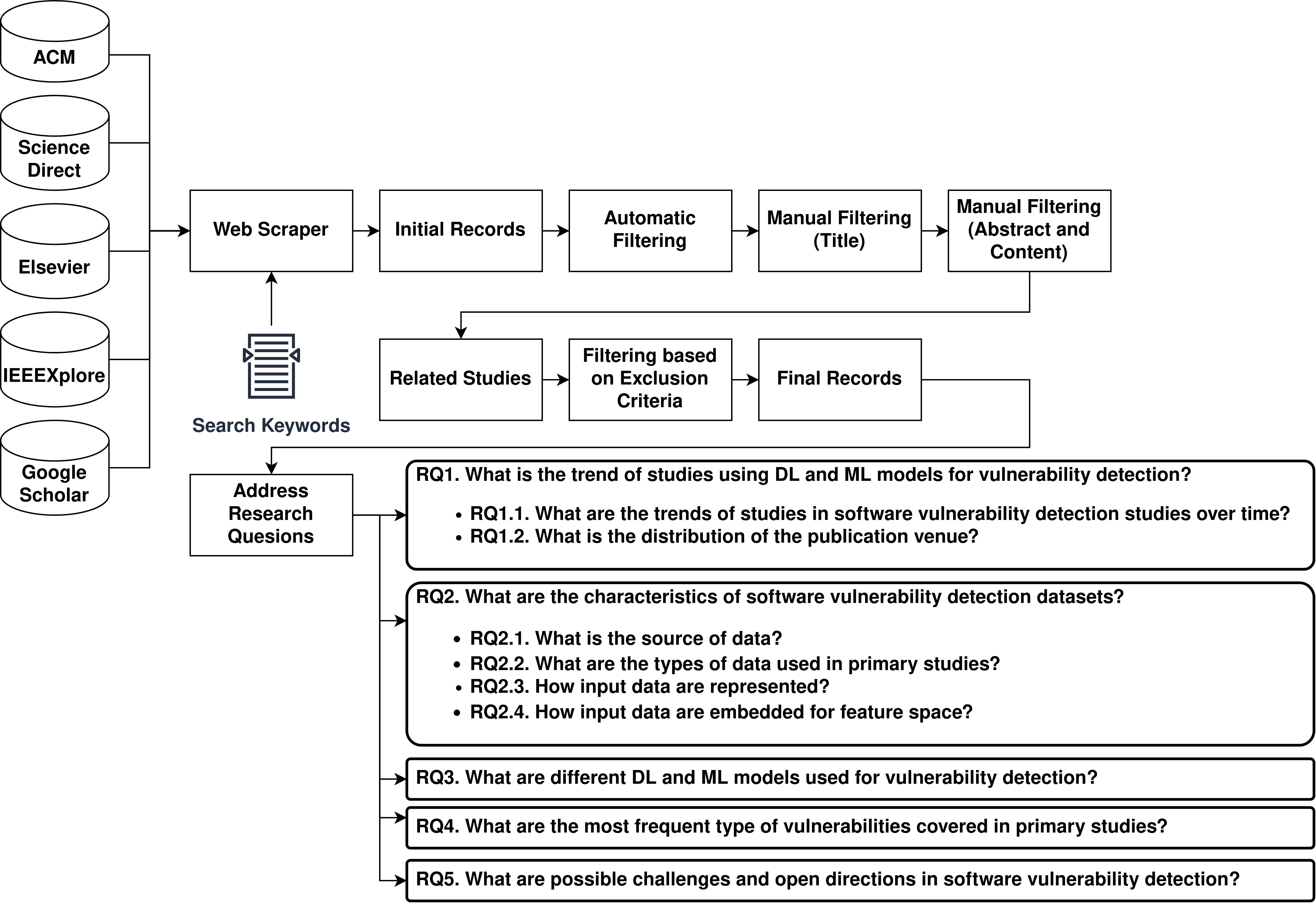}}\
     \caption{The workflow of our survey.}
    \label{vuldiag}
\end{figure}

%% file: Sections/RQ1.tex
\section{Results}


We present our analyses and findings in this section to address the research questions we devised in Section~\ref{sec:intro}. 

\subsection{RQ1. What is the trend of studies using ML/DL models for vulnerability detection?}

To comprehend the trend of publications, we examined the publication dates along with the venues in which they were presented.

\subsubsection{RQ1.1. What are the trends of studies in software vulnerability detection over time?}
Figure~\ref{fig:pub_trends} 
demonstrates the publication trend of vulnerability detection studies published in eleven years, i.e., between 2011 and 2022. It is observable that the number of publications has increased gradually over the years. There is only one publication from 2011 to 2016, and the number of publications increased to 18 in 2021. However, there is a decrease in the number of publications in 2022 compared to the previous year. We have also examined the cumulative number of publications shown in Figure \ref{fig:pub_trends}. It is noticeable that the curve fitting the distribution shows a significant increase in slope between 2018 and 2022 
suggesting that the use of ML/DL techniques for software vulnerability detection has become a prevalent trend since 2017, and a broad range of studies have utilized ML/DL models to address challenges in this field.

\input{Figures/sources/fig2}
\input{Tables/table2}
\input{Tables/table3}


\subsubsection{RQ1.2: What is the distribution of the publication venues?}
In this study, overall, we analyzed and reviewed 67 papers from various publication venues including 43 conference and symposium papers along with 24 journal papers. We have included the conference and journal acronyms and their complete names for reference in Table \ref{tbl:confLegend} and Table \ref{tbl:jourLegend}. Table \ref{tbl:venueDist}
shows the distribution of primary studies for each publication venue. 64.1\% of publications are published in conferences and symposiums while 35.8\% of papers have been published as journal papers. It is observable that MSR, IJCAI, and ECSE/FSE are the most popular venues that have the highest number of primary studies, each of which contains 4 papers. 
Meanwhile, among the journals, TDSC and TSE include the highest number of studies, i.e., 6 and 4 studies respectively.

\mybox{Answer to RQ1}{red!40}{gray!10}{
(1) The results indicate that the application of ML/DL techniques for software vulnerability detection has had a remarkable rising trend in the past few years.\\
(2) A large proportion of papers are published in recent two years, i.e., 2021 and 2022.\\
(3) MSR, IJCAI, and ECSE/FSE are the most popular conference venues. On the other hand, TDSC and TSE are the most popular journal venues.\\
}

\input{Tables/table4}

\input{Figures/sources/fig3}

%% file: Figures/sources/fig2.tex
\begin{figure}[t!]
\centering
\begin{subfigure}{.5\textwidth}
\begin{tikzpicture}[scale=0.6]
\begin{axis}  
[  
    ybar,  
    enlargelimits=0.15,  
    width=11cm,
    height=4cm,
    ylabel={Number of Publications}, 
    xlabel={Year},  
    symbolic x coords={2011, 2013, 2014,2015,2016, 2017, 2018, 2019, 2020, 2021, 2022}, 
    xtick=data,  
     nodes near coords,
    nodes near coords align={vertical},  
    ]  
\addplot coordinates {
(2022, 14)
(2021, 18)
(2020, 6)
(2019,9)
(2018,    9)
(2017,   6)
(2016,   1)
(2015,    1)
(2014,   1)
(2013,   1)
(2011,1)
};  
\end{axis}  
\end{tikzpicture}
\label{fig:pub_trends1}
\end{subfigure}%
\begin{subfigure}{.5\textwidth}
\begin{tikzpicture}[scale=0.6]
\begin{axis}  
[  
    ybar,  
    enlargelimits=0.15,  
    width=11cm,
     height=4cm,
    ylabel={Cumulative Distribution}, 
    xlabel={Year},  
    symbolic x coords={2011, 2013, 2014, 2015,2016, 2017, 2018, 2019, 2020, 2021, 2022}, 
    xtick=data,  
     nodes near coords,
    nodes near coords align={vertical},  
    ]  
\addplot coordinates {
(2011,0.01)
(2013,0.02)
(2014,0.04)
(2015,0.05)
(2016,0.07)
(2017,0.16)
(2018,0.29)
(2019,0.43)
(2020,0.52)
(2021,0.79)
(2022,1)
};  
\end{axis}  
\end{tikzpicture}
\label{fig:pub_trends2}
\end{subfigure}%
\newline
\caption{Publication trend of vulnerability detection studies.}
\label{fig:pub_trends}
\end{figure}

%% file: Tables/table2.tex
\begin{table}[t!]
\caption{Conference publication venues for manual search.}
\renewcommand{\arraystretch}{0.3}
\resizebox{\columnwidth}{!}{%
\begin{tabular}{ccl}
\toprule
No & Acronym & Full name                                                               \\
\midrule
1  & ICSE    & International Conference on Software Engineering                        \\
2  & ECSE/FSE & ACM SIGSOFT Symposium on the Foundation of Software Engineering \\
3 & ASE     & IEEE/ACM International Conference on Automated Software Engineering     \\
4 & USENIX  & USENIX                 \\
5  & OOPSLA   & Object-oriented Programming, Systems, Languages, and Applications      \\
6 & ISSTA & ACM SIGSOFT International Symposium on Software Testing and Analysis \\
7  & MSR     & IEEE Working Conference on Mining Software Repositories                 \\
8 & SANER & IEEE International Conference on Software Analysis, Evolution and Reengineering \\
9  & ISSRE   & IEEE International Symposium on Software Reliability Engineering        \\
10 & ICSME   & IEEE International Conference on Software Maintenance and Evolution     \\
11  & IJCAI   & International Joint Conferences on Artificial Intelligence Organization \\
12  & CCS     & ACM SIGSAC Conference on Computer and Communications Security           \\
13  & ICLR    & International Conference on Learning Representations                    \\
14  & NIPS    & International Conference on Neural Information Processing Systems       \\
15 & MASCOT   & Modelling, Analysis, and Simulation of Computer and Telecommunication Systems                            \\
16 & QRS     & IEEE International Conference on Software Security and Reliability      \\
17 & KDDM    & Pacific-Asia Conference on Knowledge Discovery and Data Mining          \\
18 & NDSS    & Network and Distributed Systems Security Symposium               \\
19 & ARES    & ACM International Conference on Availability, Reliability and Security  \\
20 & INFOCOM & IEEE International Workshop on Security and Privacy in Big Data         \\
21 & ICTAI   & IEEE International Conference on Tools with Artificial Intelligence     \\
22 & ICDM    & IEEE International Conference on Data Mining                            \\
23 & GLOBCOM & IEEE Global Communications Conference                                   \\
24 & TrustCom & IEEE International Conference on Trust, Security and Privacy in Computing and Communications \\
25 & DSAA & IEEE International Conference on Data Science and Advanced Analytic \\
\bottomrule
\end{tabular}}
\label{tbl:confLegend}
\end{table}

%% file: Tables/table3.tex
\begin{table}[h]
\caption{Journal publication venues for manual search.}
{%
\begin{tabular}{ccl}
\toprule
No & Journal Acronym & Full Name                                               \\
\midrule
1  & TSE             & IEEE Transaction on Software Engineering                \\
2  & TOSEM           & ACM Transaction on Software Engineering and Methodology     \\
3  & IST             & Information and Software Technology                     \\
4  & ESM             & Empirical Software Engineering                          \\
5  & JSS             & Journal of System and Software                         \\
6  & TDSC            & IEEE Transaction on Dependable and Secure Computing     \\
7  & CSJ             & Computer and Security Journal                           \\
8  & TIFS            & IEEE Transactions on Information Forensics and Security \\
9  & ISJ             & Information Sciences Journal                            \\
10 & TFS             & IEEE Transaction on Fuzzy Systems                       \\
11 & TKDE            & IEEE Transaction on Knowledge and Data Engineering \\
12 & KBS             & Knowledge-Based Systems \\
\bottomrule
\end{tabular}}
\label{tbl:jourLegend}
\end{table}

%% file: Tables/table4.tex
\begin{table}[h]
\caption{Distribution of publications based on conference and journal venues.}
\setlength\tabcolsep{2.0pt}
\begin{tabular}{lcclcc}
\toprule
Conference Venue & \# Studies & References & Journal Venue         & \# Studies & References \\
\midrule
MSR   & 4          &  \cite{chen2020machine, hoang2019deepjit, hin2022linevd, fu2022linevul}    & TDSC & 6  &   \cite{liu2020cd, zou2019mu, lin2019software, li2021sysevr, li2021vuldeelocator, zou2022mvulpreter}         \\
IJCAI            & 4          &    \cite{3172102, zhuang2020smart, duan2019vulsniper, liu2021smart}        & TSE & 4            &     \cite{dam2018automatic, dam2017automatic, chen2021neural, scandariato2014predicting}       \\
ECSE/FSE         & 4          &   \cite{zhou2017automated, li2021vulnerability, ni2022best, nguyen2022mando}      & CSJ & 2 &      \cite{jeon2021autovas, yan2021han}      \\
CCS              & 3          &       \cite{yamaguchi2013chucky, lin2017poster, perl2015vccfinder, cheng2019static}      & IST   & 2   &   \cite{shippey2019automatically, tian2020bvdetector}         \\
 ISSRE            & 3          &     \cite{zheng2021vulspg, wu2021peculiar, zeng2021gcn2defect}           & TIFS    & 2 &    \cite{wang2020combining, huang2021hunting}        \\
ICLR             & 2          &    \cite{dinella2020hoppity, le2018maximal}        & TOSEM & 2 &   \cite{cheng2021deepwukong, 3429444}         \\
ICSE             & 2          &   \cite{cao2022mvd, wang2016automatically}          & ESM  & 1 &    \cite{riom2021revisiting}        \\
OOPSLA           & 2          &    \cite{pradel2018deepbugs, 3360588}        & ISJ  & 1  & \cite{ghaffarian2021neural}   \\
NIPS             & 2          &         \cite{3327890, zhou2019devign}   & JSS  & 1  &  \cite{pascarella2019fine}          \\
  ASE         & 2          &      \cite{le2021deepcva, zhang2022reentrancy}      & TFS & 1  &  \cite{liu2019deepbalance} \\
QRS              & 1          &        \cite{li2017software}    & TKDE   & 1   &   \cite{liu2021combining}         \\
KDDM             & 1          &      \cite{nguyen2020deep}      & KBS  &  1 &   \cite{zhuang2022just}   \\
NDSS             & 1          &       \cite{li2018vuldeepecker}     &  \textbf{SUM}   & \textbf{24}  &       \\
ARES             & 1          &       \cite{kronjee2018discovering}     &     &   &      \\
INFOCOM          & 1          &     \cite{ziems2021security}       &       &   &       \\
MASCOT               & 1          &    \cite{filus2020random}          &   &  &       \\
ICTAI            & 1          &      \cite{phan2017convolutional}      &    &   &            \\
ICSME            & 1          &      \cite{sabetta2018practical}      &      &       &            \\
ICDM             & 1          &       \cite{8594942}     &    &            &    \\
GLOBCOM          & 1          &      \cite{Vuldiggerpaper}      &          &       &            \\
USENIX           & 1          &     \cite{yamaguchi2011vulnerability}    &  &           &             \\
DSAA             & 1          &     \cite{nguyen2022mando}    &  &           &             \\
ISSTA            & 1          &     \cite{cheng2022path}    &  &           &             \\
SANER            & 1          &     \cite{ding2022velvet}    &  &           &             \\
TrustCom         & 1          &     \cite{yang2022source}    &  &           &             \\
\textbf{SUM} & \textbf{43} & & & &  \\
\bottomrule
\end{tabular}
\label{tbl:venueDist}
\end{table}

%% file: Figures/sources/fig3.tex
\definecolor{airforceblue}{rgb}{0.36, 0.54, 0.66}
\definecolor{aliceblue}{rgb}{0.94, 0.97, 1.0}
\definecolor{alizarin}{rgb}{0.82, 0.1, 0.26}
\definecolor{almond}{rgb}{0.94, 0.87, 0.8}
\definecolor{amaranth}{rgb}{0.9, 0.17, 0.31}
\definecolor{amber}{rgb}{1.0, 0.75, 0.0}
\definecolor{amber(sae/ece)}{rgb}{1.0, 0.49, 0.0}
\definecolor{americanrose}{rgb}{1.0, 0.01, 0.24}
\definecolor{amethyst}{rgb}{0.6, 0.4, 0.8}
\definecolor{anti-flashwhite}{rgb}{0.95, 0.95, 0.96}
\definecolor{antiquebrass}{rgb}{0.8, 0.58, 0.46}
\definecolor{antiquefuchsia}{rgb}{0.57, 0.36, 0.51}

%% file: Sections/RQ2.tex
\subsection{RQ2. What are the characteristics of software vulnerability detection datasets?}

Data is important for building and evaluating ML/DL-based software vulnerability detection models~\cite{lin2019software, cheng2022path, lin2018cross, du2020cross, dam2017automatic}. The quality of datasets can be assessed by different factors such as the source of data, data size and scale, data types, and preprocessing steps performed on data. For example, inappropriate preprocessing (representation) on data may result in poor performance of DL models~\cite{ruospo2020evaluating}. In this section, we examine data used in vulnerability detection studies and conducted a comprehensive analysis of the steps of data source, data type, and data representation.

\subsubsection{RQ2.1. What are the sources of datasets?}
One of the main challenges in ML/DL-based software vulnerability detection is the insufficient amount of data available for training operations~\cite{chen2020machine, lin2018cross}. Consequently, there exists a gap in research on how to obtain sufficient datasets to facilitate the training of ML/DL models for software security vulnerability detection. To this end, we analyze the sources of datasets in the studied 67 primary studies. Our analysis reveals that datasets for this purpose can be broadly classified into three categories, i.e.,  \textit{Benchmark}, \textit{Collected}, and \textit{Hybrid} sources. \textit{Benchmark} contains standardized datasets used to evaluate the performance of vulnerability detection methods and techniques~\cite{sabetta2018practical, dam2018automatic, dam2017automatic, wang2016automatically, jeon2021autovas, tian2020bvdetector, liu2020cd, yamaguchi2013chucky, liu2021combining, phan2017convolutional, nguyen2020deep, hoang2019deepjit, kronjee2018discovering, zeng2021gcn2defect, yan2021han, 3429444, 8594942, 3327890}. 

Benchmark datasets for software vulnerability detection are often built from three main sources. The first source of data is collecting code snippets from open sources. This can include open-source software projects~\cite{jimenez2016empirical}, public vulnerability databases~\cite{black2017sard}, and bug repositories~\cite{bugzilla2021bugzilla}. The goal is to gather a diverse set of programs or code snippets that represent different application domains, programming languages, and vulnerability types. From the collected data, specific programs or code snippets are selected to be included in the benchmark dataset. The selection process considers factors such as program complexity, vulnerability diversity, and code quality. The aim is to create a dataset that covers a wide range of vulnerabilities and represents real-world scenarios. In some cases, benchmark datasets may include automatically generated synthetic programs~\cite{booth2013national}. These programs are typically created using code generation techniques and follow certain patterns or templates. Synthetic generation allows for the creation of large-scale datasets and can help cover a broader range of vulnerabilities systematically. Alongside synthetic programs, benchmark datasets often include real-world software applications or code snippets written by hand. These manually created cases ensure that the dataset contains realistic vulnerabilities that reflect actual coding practices. Manual creation involves identifying vulnerable points in the code, introducing appropriate weaknesses, and maintaining a balance between code quality and realism.

\textit{Collected} datasets are gathered from publicly available projects hosted on repository websites such as Github or Stack Overflow~\cite{chen2020machine, zhou2017automated, liu2019deepbalance, pradel2018deepbugs, cheng2021deepwukong, zhou2019devign}. Also, some studies use the combination of different sources for vulnerability detection to increase the external validity of their findings~\cite{chen2020machine, zhou2017automated, russell2018automated, wang2020combining, le2021deepcva, hin2022linevd, li2021vulnerability}, which refers to \textit{Hybrid} source in this work.

The distribution of dataset sources in the primary studies is illustrated in figure~\ref{fig:sources}. As we can see, 65.7\% of primary studies have utilized \textit{Benchmark} datasets for software vulnerability detection. 
The rationale behind this trend is that benchmark datasets are readily accessible to all researchers and can facilitate the reproducibility of prior studies. 
Researchers often used \textit{Collected} datasets in evaluating the proposed ML or DL-based security vulnerability detection models. According to our observation, 25.4\% of studies build vulnerability detection models using collected datasets.  There are a couple of reasons, first of all, open-source repositories like GitHub contain a vast amount of real-world code written by developers from diverse backgrounds. This data reflects the real-world coding practices, patterns, and vulnerabilities present in software projects. By analyzing such data, researchers can gain insights into the prevalent types of vulnerabilities and their occurrence frequencies in real-world software. Second, open-source repositories offer the opportunity to identify new vulnerabilities that may not be present in benchmark datasets. By analyzing diverse codebases, researchers can uncover previously unknown vulnerabilities or variations of known vulnerabilities, which helps in advancing the state-of-the-art in vulnerability detection and expanding the knowledge base of software security.
\input{Figures/sources/fig4}

The third major source of data is \textit{Hybrid} accounting for 9\% of primary studies which is the combination of different sources. Researchers often use hybrid sources for software vulnerability detection to address some of the limitations of individual data sources and to obtain more diverse and comprehensive datasets. For example, researchers may combine data from benchmark datasets with data from other sources such as Github, open-source projects, or data from commercial companies to create a hybrid dataset that is more representative of real-world scenarios. By doing so, they can improve the generalizability of their models and increase their chances of detecting a wider range of vulnerabilities. 



Table \ref{tab:benchdetail} shows the detailed distribution of benchmark data used in the primary studies. As it is observable, \textit{NVD} and \textit{SARD} is the most widely used source of data in the \textit{Benchmark} category. This is because SARD and NVD are publicly available benchmark sources to which researchers have unrestricted access. They give a lot of vulnerability data, allowing researchers to get a broad variety of vulnerabilities for their experiments and analyses. The availability of these materials promotes repeatability and collaboration among researchers in software vulnerability detection. Overall, there are 35 unique primary studies that use benchmark datasets from different sources. 

Table \ref{tab:collectdetail} shows the detailed distribution of the \textit{Collected} source of data. As shown, \textit{Github} is the most popular source of data for software vulnerability detection, accounting for 14 primary studies. 
Researchers can collect datasets from Github by crawling the platform and extracting relevant code repositories or by using Github's API to access data programmatically. One advantage of using GitHub as a source of data is that it provides access to real-world code written by developers, which can be used to train and test vulnerability detection models. This data can be particularly useful for detecting new and emerging vulnerabilities that may not be covered by benchmark datasets. Jira, CVE, and Bugzilla come after with 2 primary studies for each. Overall, there are 17 unique primary studies that use collected sources of data for software vulnerability detection.

\input{Tables/table5}

\subsubsection{RQ2.2. What are the types of software vulnerability detection datasets used in prior studies?}

When it comes to detecting software vulnerabilities, datasets can have varying data types, e.g., existing software vulnerability detection models can find vulnerabilities in source code or commits. It is crucial to carefully examine the data types, as they require different preprocessing techniques and must be represented differently when using ML/DL models. Additionally, distinct data types necessitate different architectural approaches for ML/DL models. This section provides an overview of the various data types and their distributions. We classified the data types of employed datasets into three broad categories, i.e., \textit{code-based}, \textit{repository-based}, and \textit{hybrid}.

Figure \ref{fig:datasetsDist} shows the distribution of the data types in primary studies. We can observe that the majority of primary studies (70.1\%) primarily focus on analyzing the source code for software vulnerability detection. This indicates the significance of code-level analysis in identifying vulnerabilities. The utilization of repository-level data, such as commit history and change logs, is also prominent, representing a substantial portion (25.4\%) of primary studies. This suggests that repository-level information is considered valuable in vulnerability detection. Additionally, a smaller portion (4.5\%) of the studies adopt a hybrid approach, combining both code-level analysis and repository-level information. These techniques leverage the strengths of both data sources to improve the accuracy and effectiveness of vulnerability detection.

\begin{figure}[t!]
  \centering
\begin{tikzpicture}[scale=0.6]
\pie{70.1/Code,
    25.4/Repository,
    4.5/Hybrid}
\end{tikzpicture}
  \caption{The type of datasets in primary studies.}
  \label{fig:datasetsDist}
\end{figure}%

\input{Tables/table6}

Table \ref{datatypesdist} elaborates the detailed data types categories used in primary studies. The table shows that 42 primary studies used code-based category and the major data type of this category is \textit{Source code} \cite{zheng2021vulspg, duan2019vulsniper, yamaguchi2011vulnerability, li2021vulnerability, zou2019mu, li2021vuldeelocator, filus2020random, li2021sysevr, lin2019software, li2017software, zhuang2020smart, ziems2021security}. 
\textit{Binary code} is the second major data type in code-based category \cite{phan2017convolutional, huang2021hunting, yan2021han, nguyen2020deep} accounting for 5 primary studies. 
Regarding the \textit{Repository} based category, 13 primary studies focused on extracting useful information and patterns by crawling different artifacts from software repositories from open source. The major artifact is \textit{Code change} accounting for 8 primary studies~\cite{chen2021neural, fu2022linevul, hin2022linevd, 3360588, dinella2020hoppity, zhou2019devign, pradel2018deepbugs, liu2019deepbalance}, and \textit{Commit} comes as the second with 5 primary studies. The last category of data types is \textit{Hybrid} where the studies used a combination of different data types for software security vulnerability detection, accounting for 7 primary studies. As can be seen, \textit{Source code+Code change} is the most dominant data type combination~\cite{wang2020combining, le2021deepcva, cheng2021deepwukong}.
\input{Figures/sources/fig5}



\subsubsection{RQ2.3. How input data are represented?}
As noted in earlier sections, research studies focused on software vulnerability detection rely on diverse sources of data and data types. This heterogeneity necessitates the use of varied representation techniques, which in turn requires different architectural approaches and design assumptions for ML/DL models.
We classified the input representation of employed datasets into five broad categories, i.e., \textit{Graph-based}, \textit{Tree-based}, \textit{Token-based}, \textit{Metric-based} and \textit{Hybrid}. Figure \ref{fig:representationDist} shows the distribution of different input representations used in primary studies. From the pie chart, we can observe that the most popular input representation is the use of \textit{Graph/Tree-based representation}, accounting for the largest slice (i.e., 38.8\% for Graph-based and 16.4\% for Tree-based). \textit{Token-based representation} follows closely, representing a substantial portion (29.9\%) of primary studies. \textit{Hybrid representation} combines multiple representations or approaches, which makes up a smaller portion (10.4\%). Finally, the use of \textit{Commit Metrics} in vulnerability detection has the smallest portion (4.5\%). In the following paragraphs, we elaborate on each category in detail. 
\noindent \textbf{Graph/Tree-based representation}~\cite{jeon2021autovas, tian2020bvdetector, liu2021combining, phan2017convolutional,
cheng2021deepwukong, zhou2019devign, kronjee2018discovering, zeng2021gcn2defect,
huang2021hunting, 3360588, zou2019mu, cao2022mvd,
ghaffarian2021neural, wu2021peculiar, zhuang2020smart, li2021sysevr,
zou2019mu, duan2019vulsniper, zheng2021vulspg, cheng2022path,
liu2021smart, ding2022velvet, yang2022source, zou2022mvulpreter,
 nguyen2022mando, nguyen2022mando2}: 
allows for the detection of complex patterns and relationships between different code elements. By representing source code as a graph or tree, it becomes possible to capture not only the syntax and structure of the code but also its semantics, control flow, and data flow. There are many graph/tree-based representation techniques like Abstract Syntax Trees (AST) \cite{mao2020explainable, yamaguchi2012generalized, dinella2020hoppity, lin2017poster, lin2019software, li2021vuldeelocator} and Code Property Graph (CPG) \cite{zhou2019devign, ghaffarian2021neural, duan2019vulsniper} used to transform source code into AST and CPG representations.

\begin{figure}{t!}
  \centering
    \begin{tikzpicture}[scale=0.6]
    \pie{29.9/Token,
        38.8/Graph,
        16.4/Tree,
        10.4/Hybrid,
        4.5/Commit Metrics
        }
    \end{tikzpicture}
  \caption{Different representation used by primary studies.}
  \label{fig:representationDist}
\end{figure}%

\noindent \textbf{Token-based representation}~\cite{aivatoglou2021tree, zhou2017automated, russell2018automated, dam2018automatic, yamaguchi2013chucky, du2020cross, nguyen2020deep, liu2019deepbalance, pradel2018deepbugs, le2021deepcva, hoang2019deepjit}: treat the source code as string token sequences and then transforms source code into tokens vectors. The input data is first split into a sequence of tokens, which are then converted into numerical vectors that can be processed by machine learning algorithms. Tokenization involves breaking down a string of text or source code into smaller units, or tokens, which can then be used as the basis for further analysis. In the case of source code, tokens might include keywords, operators, variables, and other elements of the programming language syntax.

\noindent \textbf{Commit Metrics}~\cite{ni2022best, Vuldiggerpaper, pascarella2019fine}: 
leverages the metrics extracted from commits to represent code commits. Features derived from commits, such as the size of code changes, the number of modified lines, the complexity of the changes, or the programming language used, can be used as inputs to train ML/DL models. These models can then learn patterns and relationships between commit characteristics and the presence of vulnerabilities, enabling automated detection based on new commits.

\noindent \textbf{Hybrid representation}~\cite{russell2018automated, wang2020combining, le2021deepcva, cheng2021deepwukong, sabetta2018practical, hoang2019deepjit, zhou2017automated, chen2020machine}: uses a combination of different representations for software security vulnerability detection. Combining different representations of input data can lead to a more comprehensive and richer input representation of source code, which can improve the performance of vulnerability detection models in tasks such as prediction or detection. Combining different representations such as token-based representations and graph-based representations can help capture both the syntax and semantics of the code, as well as the relationships between different components of the code.


Table \ref{tbl:represenation_dist_tbl} shows the representation techniques distributed by different artifacts used by ML/DL models. It is observable that \textit{Graph/Tree-based representation} is the most dominant technique used by primary studies, accounting for 32 unique primary studies in total. These studies represent the input to ML/DL models via \textit{Source code as a graph}, \textit{Source code as a tree}, and \textit{Binary code as a graph}. \textit{Source code as a graph} is the major representation technique used by primary studies \cite{jeon2021autovas, tian2020bvdetector, liu2021combining, phan2017convolutional,
cheng2021deepwukong, zhou2019devign, kronjee2018discovering, zeng2021gcn2defect,
huang2021hunting, 3360588, zou2019mu, cao2022mvd,
ghaffarian2021neural, wu2021peculiar, zhuang2020smart, li2021sysevr,
zou2019mu, duan2019vulsniper, zheng2021vulspg, cheng2022path,
liu2021smart, ding2022velvet, yang2022source, zou2022mvulpreter,
nguyen2022mando, nguyen2022mando2} accounting for 22 studies. \textit{Source code as a tree}~\cite{dam2017automatic,
 shippey2019automatically,
 wang2016automatically,
 liu2020cd,
 wang2020combining,
 dinella2020hoppity,
 lin2017poster,
 li2017software,
 lin2019software,
 li2021vuldeelocator,
 zhuang2022just} is the second major representation technique accounting for 9 primary studies. Some researchers used \textit{Binary code as a graph}~\cite{phan2017convolutional, huang2021hunting} to build binary-level vulnerability detection models, accounting for 2 primary studies. 
There are 14 primary studies that used \textit{Token-based representation}, in which 10 primary studies represented source code as a token sequence, three primary studies modeled binary code as a token, and one study represented text as token sequences~\cite{zhou2017automated}. 
{\textit{Hybrid} representation has 5 different types accounting for 8 primary studies. \textit{Token Sequence+Commit Metrics} is the major artifact used to enhance the input representation in software vulnerability detection, accounting for 4 primary studies. It combines information from the token sequence of the code and additional metrics derived from software commits. This approach leverages both the structural and historical aspects of the code to provide a more comprehensive representation for building vulnerability detection models.}

\textit{Commits} is the fourth least input representation used by 3 primary studies. In this representation, commit characteristics are used to build software vulnerability detection models.

\input{Tables/table7}

\input{Figures/sources/fig6}

Figure \ref{typeDist} shows the distribution of data type representation in software vulnerability detection studies over time. As shown in the figure, \textit{Graph-based} representation shows a substantial presence compared to other input representation techniques. There are a couple of reasons for this trend. First, graphs provide a natural and intuitive way to represent the structural relationships within the source code. By modeling the code as a graph, the relationships between functions, classes, methods, and variables can be captured effectively. This allows vulnerability detection algorithms to analyze the code at a higher level of abstraction and capture complex dependencies and interactions between code elements. Second, graph-based representations enable a better understanding of the context in which vulnerabilities may exist. By considering the surrounding code structure and dependencies, graph-based approaches can capture the flow of information and identify potential paths that can lead to vulnerabilities. This contextual understanding helps in identifying code patterns, control flow paths, and data dependencies that may introduce security vulnerabilities. \textit{Token-based} representation has also gained popularity, with a peak occurrence in 2021. This is because it provides a fine-grained representation of the code. It simplifies the code analysis process by reducing the complexity of the code to a sequence of tokens, making it easier to apply traditional natural language processing techniques or ML models. It is also easily applicable to a wide range of programming languages. While the tokens themselves may differ across languages, the concept of breaking the code into discrete units remains the same. This versatility allows vulnerability detection techniques based on token representation to be applied to different programming languages and codebases, which further increases the external validity of vulnerability detection models. However, there is a slight decline in 2022, indicating potential shifts or diversification in the selection of input representations. \textit{Hybrid} representation is gained attention since 2021, which suggests that combining different representations is favored by researchers in software vulnerability detection, potentially due to the complementary benefits provided by multiple representations.

\subsubsection{RQ2.4. How input data are embedded for feature space?}
In the previous section, we discussed various representation techniques, and in this section, we further look at embedding methods that can transform these representations into inputs that can be understood by ML/DL models. 
The representation techniques are in a human-readable format, they cannot be directly interpreted by machines. Therefore, researchers use different embedding techniques to convert these representations into a numeric format. We discuss the embedding techniques in the following paragraphs based on the distribution shown in Figure~\ref{fig:embedding_dist}. The figure illustrates the distribution of feature embedding techniques used in primary studies. The chart shows the following categories and their corresponding percentages: Word2vec (25.4\%), Graph embedding (25.4\%), Token vector embedding (11.9\%), Others (16.4\%), Hybrid (7.5\%), One hot embedding (6.0\%), Code token embedding (4.5\%), and N-gram features (3.0\%).

\input{Figures/sources/fig7}

\noindent \textbf{Word2vec}~\cite{chen2020machine, sabetta2018practical, zhou2017automated, jeon2021autovas, yamaguchi2013chucky, du2020cross, lin2018cross, liu2019deepbalance, pradel2018deepbugs, zhou2019devign, yan2021han, 3360588, 3429444, 8594942, zou2019mu, scandariato2014predicting, riom2021revisiting, lin2019software, perl2015vccfinder, li2021vulnerability}: is one of the most widely-used embedding techniques for source code embedding in the examined papers, accounting for 25.4\% of primary studies. This can be because it has been shown to be effective in capturing the semantics and relationships between different code components. Word2vec can be trained on code corpus to learn embeddings for different code components, such as variables, functions, and operators. By considering the context in which these components appear, Word2vec can capture the semantic relationships between them. Furthermore, Word2vec is a computationally efficient and scalable technique, which can be trained on large code corpora. This is important for source code embedding, as the code corpus can be much larger than the text corpus typically used in natural language processing.

\noindent \textbf{Graph embedding}~\cite{liu2021combining, wang2020combining, phan2017convolutional, zeng2021gcn2defect, dinella2020hoppity, cao2022mvd, ghaffarian2021neural, wu2021peculiar, zhuang2020smart, duan2019vulsniper, zheng2021vulspg}: is another widely-used embedding technique among the primary studies, accounting for 25.4\% of primary studies, which is mostly used by graph neural networks. 
This can be because it can capture the structural relationships between different code components, such as functions, classes, and methods. In contrast to token-based representations or sequence-based representations, graph embedding can explicitly represent the connections and dependencies between different code components. In a graph-based representation, code components are represented as nodes, and the relationships between them are represented as edges. This allows for a more fine-grained representation of the code structure. 

\noindent \textbf{Token vector embedding}~ \cite{dam2018automatic, dam2017automatic, liu2020cd, wang2016automatically, hoang2019deepjit, mao2020explainable, yamaguchi2012generalized, hin2022linevd, fu2022linevul, chen2021neural, lin2017poster, yamaguchi2011vulnerability}: is also a popular technique used by primary studies accounting for 11.9\% of examined papers. In this technique, input is converted into a sequence of tokens and each token is transformed into a numeric value. Then, these values are fed into ML/DL models for further computations. 

\noindent \textbf{One hot embedding}~\cite{nguyen2020deep, hoang2019deepjit, 3327890, le2018maximal}: is a typical way for encoding categorical data, in which each category is represented by a binary vector of zeros and ones. This method can also be used to encode source code for vulnerability detection, which accounts for 6\% of studies.

\noindent \textbf{Code token embedding}~\cite{dam2018automatic, dam2017automatic, liu2020cd}: is used to represent source code tokens as dense vectors in a continuous vector space. Code token embedding captures the semantic and syntactic links between tokens by transferring them to a lower-dimensional vector space, as opposed to one hot encoding, which represents each token as a sparse binary vector.

\noindent \textbf{N-gram features}~\cite{shippey2019automatically, le2021deepcva}: is a method of expressing code snippets as fixed-length dense vectors, each vector representing an n-gram of tokens. N-grams are sequences of \verb|n| tokens, such as words or letters, that capture local context and interdependence between neighboring tokens. We observed that 3\% of primary studies use N-gram features for embedding.

\noindent \textbf{Hybrid} ~\cite{jeon2021autovas,
cheng2021deepwukong,
zhou2019devign,
3360588,
li2021vulnerability}: We find that 7.5\% of primary studies use multiple embedding techniques to convert inputs to ML/DL models. Different embedding techniques capture different aspects of the data. By combining multiple techniques, researchers can leverage the complementary information provided by each technique. For example, some embedding techniques may focus on syntax, while others may capture semantic or contextual information.

\noindent \textbf{Others}~\cite{sabetta2018practical,
kronjee2018discovering,
pascarella2019fine,
huang2021hunting,
ziems2021security,
filus2020random,
Vuldiggerpaper,
zhuang2022just,
cheng2022path,
zhang2022reentrancy,
ni2022best}: The remaining 16.4\% that emerge seldom and do not belong to any group are classified as \textit{Others}. For example, Zhang et al.~\cite{zhang2022reentrancy} customizes the graphCodeBERT~\cite{guo2020graphcodebert} to propose a graph-guided masked attention mechanism for vulnerability detection in which it captures variable dependency relationships and integrates the graph structure into the Transformer model.


\mybox{Answer to RQ2}{red!40}{gray!10}{
(1) 65.7\% of primary studies use \textit{benchmark data} for software vulnerability detection. This can be because benchmark datasets are readily accessible to all researchers and can facilitate the reproducibility of studies. 


(2) The most common data type among the examined vulnerability detection studies is \textit{Code-based data type}, accounting for 47 studies. In this category, \textit{Source code} is the most prominent sub-type accounting for 42 studies. 


(3) \textit{Graph-based} and \textit{Token-based} input representations are the most popular input representation techniques used by primary studies accounting for 38.8\% and 29.9\% of primary studies respectively. 

(4) \textit{Graph embedding} and \textit{Word2vec} are the two most widely used embedding techniques used in primary studies accounting for 25.4\% of studies respectively. 
}

%% file: Figures/sources/fig4.tex
\begin{figure}[t!]
\centering
\begin{tikzpicture}[scale=0.5]
\pie{65.7/Benchmark,
    25.4/Collected,
    9/Hybrid}
\end{tikzpicture}
\caption{The source of the datasets used in primary study papers.}
\label{fig:sources}
\end{figure}

%% file: Tables/table5.tex
\begin{table}[t]
\caption{Detailed distribution of benchmark sources.}
\begin{tabular}{llcp{0.55\linewidth}}
\toprule
No & Source    & \# Studies & References \\
\midrule
1  & SARD    & 17  &  \cite{jeon2021autovas,
 tian2020bvdetector,
 liu2020cd,
 wang2020combining,
 cheng2021deepwukong,
 zou2019mu,
 cao2022mvd,
 ziems2021security,
 lin2019software,
 li2021sysevr,
 filus2020random,
 li2021vuldeelocator,
 zou2019mu,
 li2021vulnerability,
 duan2019vulsniper,
 zheng2021vulspg,
 yang2022source}  \\
2  & NVD    & 13  &  \cite{jeon2021autovas,
 liu2020cd,
 wang2020combining,
 le2021deepcva,
 kronjee2018discovering,
 hin2022linevd,
 zou2019mu,
 cao2022mvd,
 li2021sysevr,
 filus2020random,
 li2021vuldeelocator,
 zou2019mu,
 zheng2021vulspg}  \\
3  & ESC and VSC & 3  &  \cite{liu2021combining, zhuang2020smart, liu2021smart}  \\
4  & SmartBugs Wild & 3  &  \cite{wu2021peculiar, nguyen2022mando, nguyen2022mando2}  \\
5  & Juliet test suit & 3  &  \cite{yan2021han, li2021vulnerability, ding2022velvet}  \\
6  & SmartBugs & 2 &  \cite{nguyen2022mando, nguyen2022mando2}  \\
7  & PROMISE    & 2 &  \cite{wang2016automatically, zeng2021gcn2defect}  \\
8  & D2A        & 2   & \cite{cheng2022path, ding2022velvet} \\
9  & NDSS    & 2  &  \cite{le2018maximal, nguyen2020deep}  \\ 
10  & NIST    & 1  &  \cite{3327890}  \\
11  & OWASP    & 1  &  \cite{3327890}  \\
12 & SAMATE    & 1      &  \cite{kronjee2018discovering}  \\
13  & Mozilla Firefox projects    & 1  &  \cite{Vuldiggerpaper}  \\
14  & ICLR2019  & 1  &  \cite{Vuldiggerpaper}  \\
15  & FQ  & 1  &  \cite{cheng2022path}  \\
16  & Bugs Wild Dataset  & 1  &  \cite{zhang2022reentrancy}  \\
17 & Others    & 1  &  \cite{nguyen2020deep}  \\
\hline
  - & \textbf{SUM}   & \textbf{53 (35)} & - \\
\bottomrule
\end{tabular}
\label{tab:benchdetail}
\end{table}

\begin{table}[t!]
\caption{Detailed distribution of Collected sources.}
\begin{tabular}{llcc}
\toprule
No & Source    &  \# Studies & References \\
\midrule
1  & Github    & 14  &  \cite{chen2020machine, zhou2017automated, liu2019deepbalance, pradel2018deepbugs, cheng2021deepwukong, zhou2019devign, pascarella2019fine, dinella2020hoppity, 3360588, fu2022linevul, chen2021neural, perl2015vccfinder, riom2021revisiting, ni2022best}  \\
2 & Jira & 2 & \cite{chen2020machine, zhou2017automated}  \\
3 & CVE & 2 & \cite{chen2020machine, zou2022mvulpreter} \\
4 & Bugzilla & 2 & \cite{chen2020machine, zhou2017automated} \\
6 & Others & 2& \cite{huang2021hunting, hoang2019deepjit} \\
\hline
  - & \textbf{SUM}   & \textbf{22 (17)} & - \\
\bottomrule
\end{tabular}
\label{tab:collectdetail}
\end{table}

%% file: Tables/table6.tex
\begin{table}[t!]
\caption{Data types of datasets involved in primary studies.}
\scalebox{0.9}{
\begin{tabular}{lp{0.3\linewidth}llp{0.35\linewidth}}
\toprule
Category & Data Type             & \#Studies & Total & References\\
\midrule
\multirow{1}{*}{Code-based}                & Source code             & 42 & \multirow{1}{*}{47} & \cite{dam2018automatic,
dam2017automatic,
shippey2019automatically,
wang2016automatically,
jeon2021autovas,
tian2020bvdetector,
liu2020cd,
yamaguchi2013chucky,
liu2021combining,
kronjee2018discovering,
3172102,
zeng2021gcn2defect,
3429444,
8594942,
3327890,
zou2019mu,
cao2022mvd,
ghaffarian2021neural,
wu2021peculiar,
lin2017poster,
scandariato2014predicting,
ziems2021security,
zhuang2020smart,
li2017software,
lin2019software,
li2021sysevr,
filus2020random,
li2021vuldeelocator,
li2018vuldeepecker,
li2021vulnerability,
yamaguchi2011vulnerability,
duan2019vulsniper,
zheng2021vulspg,
zhuang2022just,
cheng2022path,
liu2021smart,
ding2022velvet,
zhang2022reentrancy,
yang2022source,
zou2022mvulpreter,
nguyen2022mando,
nguyen2022mando2} \\
         & Binary code             & 5         &   &  \cite{phan2017convolutional,
nguyen2020deep,
yan2021han,
huang2021hunting,
le2018maximal}  \\
\hline
\multirow{1}{*}{Repository-based} & Code change             & 8  & \multirow{1}{*}{13} & \cite{liu2019deepbalance, pradel2018deepbugs, zhou2019devign, dinella2020hoppity, 3360588, hin2022linevd, fu2022linevul, chen2021neural} \\
         & Commit                & 5         &      & \cite{pascarella2019fine, perl2015vccfinder, riom2021revisiting, Vuldiggerpaper, ni2022best} \\
\hline
\multirow{1}{*}{Hybrid} & Source Code+Code change & 3  & \multirow{1}{*}{7} & \cite{wang2020combining, le2021deepcva, cheng2021deepwukong} \\
 & Commits+Code Change                & 2         &      & \cite{sabetta2018practical, hoang2019deepjit}  \\
& Commits+Bug reports                & 1         &     &   \cite{zhou2017automated} \\
& Bug report+Commits+Emails & 1         &     &   \cite{chen2020machine} \\
\bottomrule
\textbf{SUM} & - &      -    &   67(67)   &  - \\
\bottomrule
\end{tabular}}
\label{datatypesdist}
\end{table}

%% file: Figures/sources/fig5.tex

%% file: Tables/table7.tex
\begin{table}[t!]
\caption{Distribution of input representations in primary studies.}
\setlength{\tabcolsep}{3pt}
\scalebox{0.9}{
\begin{tabular}{p{0.13\linewidth}lllp{0.35\linewidth}}
\toprule
Category                                    & Artifact                      & \#Studies & Total  & References \\
\midrule
\multirow{1}{*}{Graph/Tree-based} & Source code as a graph                 & 22        & \multirow{1}{*}{32(32)}   &    \cite{jeon2021autovas,
tian2020bvdetector,
liu2021combining,
kronjee2018discovering,
zeng2021gcn2defect,
3360588,
zou2019mu,
cao2022mvd,
ghaffarian2021neural,
wu2021peculiar,
zhuang2020smart,
li2021sysevr,
li2018vuldeepecker,
duan2019vulsniper,
zheng2021vulspg,
cheng2022path,
liu2021smart,
ding2022velvet,
yang2022source,
zou2022mvulpreter,
nguyen2022mando,
nguyen2022mando2}              \\
                                            & Source code as a tree                  & 9        &    &     \cite{dam2017automatic,
shippey2019automatically,
wang2016automatically,
liu2020cd,
lin2017poster,
li2017software,
lin2019software,
li2021vuldeelocator,
zhuang2022just}             \\
                                            & Binary code as graph             & 2         &    &     \cite{phan2017convolutional,
huang2021hunting}             \\
                                            \hline
\multirow{1}{*}{Token-based} & Source code as a token                 & 10        & \multirow{1}{*}{14(14)}   & \cite{dam2018automatic,
 yamaguchi2013chucky,
 hoang2019deepjit,
 3429444,
 8594942,
 3327890,
 scandariato2014predicting,
 ziems2021security,
 filus2020random,
 yamaguchi2011vulnerability}                 \\
                                            & Binary code as a token             & 3         &    &      \cite{nguyen2020deep, yan2021han, le2018maximal}            \\
                                            & Text as a token                 & 1         &    &     \cite{zhou2017automated}             \\
\hline
\multirow{1}{*}{Hybrid} & Token Sequence+Commit Metrics & 4 & \multirow{1}{*}{8(8)}                 &      \cite{chen2020machine, perl2015vccfinder, riom2021revisiting, ni2022best}            \\
                                            & Token Sequence+Term frequency & 1         &    &      \cite{sabetta2018practical}            \\
                                            & Token Sequence+ Graph         & 1         &    &      \cite{fu2022linevul}            \\
                                            & Graph+Tree+Token Sequence     & 1         &    &      \cite{li2021vulnerability}            \\
                                            & Token+Tree     & 1         &    &      \cite{zhang2022reentrancy}            \\
\hline
\multirow{1}{*}{Commits} & Commit Metrics & 3 & \multirow{1}{*}{3}                 &      \cite{pascarella2019fine,
Vuldiggerpaper,
ni2022best}            \\
                                            \hline
\multirow{1}{*}{SUM} & - & - & \multirow{1}{*}{57(55)}                 &      \cite{pascarella2019fine,
Vuldiggerpaper,
ni2022best}            \\         
\bottomrule
\end{tabular}
}
\label{tbl:represenation_dist_tbl}
\end{table}

%% file: Figures/sources/fig6.tex

\definecolor{ultrapink}{rgb}{1.0, 0.44, 1.0}
\pgfplotstableread{
Date  Commit Graph 	Hybrid 	Token 	Tree
2011 	0 	0 	0 	1 	0
2013 	0 	0 	0 	1 	0
2014	0 	0 	0 	1 	0
2015	0 	0 	1 	0 	0
2016 	0 	0 	0 	0 	1
2017 	1 	1 	0 	2 	2
2018 	0 	3 	1 	4 	1
2019 	1 	3 	0 	3 	2
2020 	0 	2 	1 	2 	1
2021 	0 	10 	2 	4 	2
2022 	1 	7 	2 	2 	2
}\testdata
\begin{figure}[t!]
\centering
\renewcommand{\arraystretch}{1}
\resizebox{0.6\columnwidth}{!}{%
    \begin{tikzpicture}[scale=0.6]
    \begin{axis}[
        ybar stacked,
    	bar width=15pt,
        ymin=0,
        ymax=20,
        xtick=data,
        legend style={at={(1.01,0.54)},anchor=west},
        reverse legend=true,
        xticklabels={ {2011}, {2013}, {2014}, {2015}, {2016}, {2017}, {2018}, {2019},
       {2020}, {2021}, {2022}},
        xticklabel style={rotate=25,anchor=east},
    ]
    \addplot + [ nodes near coords, fill=green!80]      table    [y=Commit, meta=Date, x expr=\coordindex] {\testdata};
    \addplot + [ nodes near coords, fill=cyan!80]      table    [y=Graph, meta=Date, x expr=\coordindex] {\testdata};
    \addplot + [nodes near coords, fill=blue!60]         table    [y=Hybrid, meta=Date, x expr=\coordindex] {\testdata};
    \addplot + [nodes near coords, fill=red!60]         table    [y=Token, meta=Date, x expr=\coordindex] {\testdata};
    \addplot + [nodes near coords, fill=ultrapink!60] table    [y=Tree, meta=Date, x expr=\coordindex] {\testdata};
\legend{Commits,
Graph,
Hybrid,
Token,
Tree
}
\end{axis}
\end{tikzpicture}}
\caption{Distribution of data type representations in software vulnerability detection studies over time.}
\label{typeDist}
\end{figure}
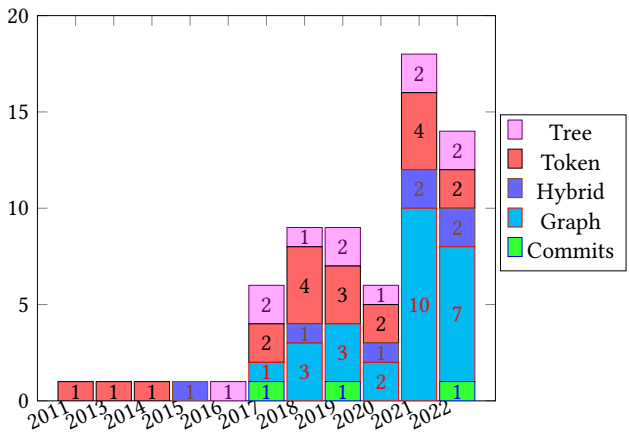

%% file: Figures/sources/fig7.tex
\begin{figure}[t!]
    \centering
    \begin{tikzpicture}[scale=0.6]
    \pie{25.4/Word2vec,
        25.4/Graph embedding,
        11.9/Token vector embedding,
        16.4/Others,
        7.5/Hybrid,
        6.0/One hot embedding,
        4.5/Code token embedding,
        3.0/N-gram features
        }
    \end{tikzpicture}
     \vspace{-0.1in}
     \caption{Different feature embedding techniques used in primary studies.}
    \label{fig:embedding_dist}
\end{figure}
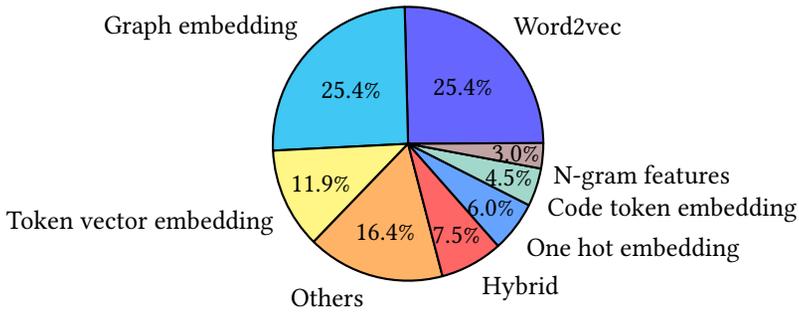

%% file: Sections/RQ3.tex
\subsection{RQ3. What are the ML/DL models used for software vulnerability detection?}

{In this section, we provide detailed information about the various ML/DL models utilized for software vulnerability detection. Initially, we present an analysis of the usage distribution of ML/DL models based on primary studies. Subsequently, we delve into the distribution of the usage of DL models used in primary studies over time. However, we have not extensively analyzed the distribution of ML models since their prevalence is relatively small compared to DL models. Nonetheless, we provide a comprehensive list of classic ML models that have been commonly employed in primary studies.}


\begin{figure}[t!]
\centering
\begin{tikzpicture}[scale=0.5]
    \pie{79.1/DL,
    16.4/ML,
    3.0/DM,
    1.5/LM
    }
\end{tikzpicture}
\caption{Distribution of models used in primary studies. DM stands for Distance Measure and LM stands for Language Model.}
\label{fig:models}
\end{figure}
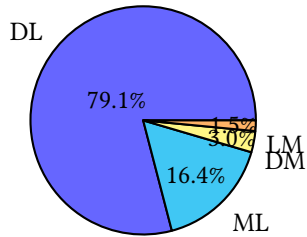
 

From Figure \ref{fig:models}, it is observable that 79.1\% of studies are using DL models for software vulnerability detection \cite{russell2018automated, dam2018automatic, dam2017automatic, wang2016automatically, jeon2021autovas, an2020avdhram, tian2020bvdetector, liu2020cd, liu2021combining, wang2020combining, phan2017convolutional, lin2018cross, liu2019deepbalance, pradel2018deepbugs, le2021deepcva, hoang2019deepjit, cheng2021deepwukong, zhou2019devign} while merely 16.4\% of studies use classic ML models \cite{nguyen2020deep, kronjee2018discovering, pascarella2019fine, yamaguchi2012generalized, scandariato2014predicting, riom2021revisiting, perl2015vccfinder, yamaguchi2011vulnerability}. Also, a limited number of studies use Language models denoted as \textit{LM} \cite{shippey2019automatically, pang2015predicting} and Distance Measures denoted as \textit{DM} \cite{yamaguchi2013chucky, huang2021hunting}. 

\definecolor{yellow-green}{rgb}{0.6, 0.8, 0.2}
\definecolor{verdigris}{rgb}{0.26, 0.7, 0.68}
\definecolor{vermilion}{rgb}{0.89, 0.26, 0.2}
\definecolor{ultrapink}{rgb}{1.0, 0.44, 1.0}
\definecolor{orange}{rgb}{1,0.5,0}

\pgfplotstableread{
DLMODELS Attention CNN GNN General Hybrid RNN Transformers
2016 	0 	0 	0 	1 	0 	0 	0
2017 	0 	1 	1 	1 	0 	1 	0
2018 	0 	1 	0 	1 	1 	3 	0
2019 	2 	1 	1 	0 	1 	2 	0
2020 	0 	0 	2 	0 	1 	6 	0
2021 	1 	0 	11 	0 	2 	8 	1
2022 	3 	0 	5 	0 	0 	3 	3
}\testdata
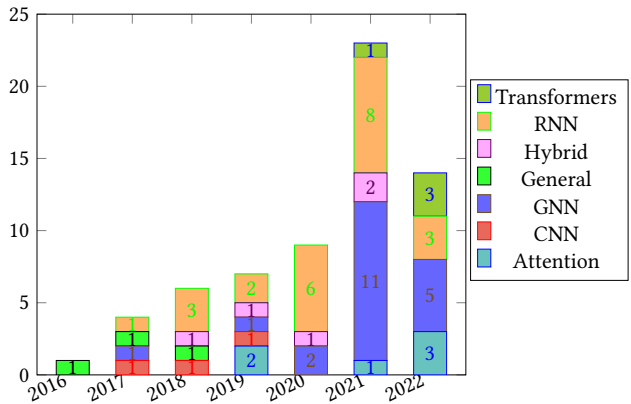
\begin{figure}[h]
\centering
\renewcommand{\arraystretch}{1}
\resizebox{0.6\columnwidth}{!}{%
    \begin{tikzpicture}[scale=0.6]
    \begin{axis}[
        ybar stacked,
    	bar width=15pt,
        ymin=0,
        ymax=25,
        xtick=data,
        legend style={at={(1.01,0.54)},anchor=west},
        reverse legend=true,
        xticklabels={ {2016}, {2017}, {2018}, {2019},
       {2020}, {2021}, {2022}},
        xticklabel style={rotate=25,anchor=east},
    ]
    \addplot + [ nodes near coords, fill=verdigris!80]      table    [y=Attention, meta=DLMODELS, x expr=\coordindex] {\testdata};
    \addplot + [ nodes near coords, fill=vermilion!80]      table    [y=CNN, meta=DLMODELS, x expr=\coordindex] {\testdata};
    \addplot + [nodes near coords, fill=blue!60]        table    [y=GNN, meta=DLMODELS, x expr=\coordindex] {\testdata};
    \addplot + [ nodes near coords, fill=green!80]      table    [y=General, meta=DLMODELS, x expr=\coordindex] {\testdata};
    \addplot + [nodes near coords, fill=ultrapink!60]         table    [y=Hybrid, meta=DLMODELS, x expr=\coordindex] {\testdata};
    \addplot + [nodes near coords, fill=orange!60] table    [y=RNN, meta=DLMODELS, x expr=\coordindex] {\testdata};
    \addplot + [nodes near coords, fill=yellow-green] table    [y=Transformers, meta=DLMODELS, x expr=\coordindex] {\testdata};
\legend{Attention,
CNN,
GNN,
General,
Hybrid,
RNN,
Transformers
}
\end{axis}
\end{tikzpicture}}
\caption{Trend of DL models over time.}
\newcommand{\squeezeup}{\vspace{-10.5mm}}
\label{fig:dl_over_time}
\end{figure}

The graph in Figure \ref{fig:dl_over_time} illustrates the usage trend of DL models in detecting software vulnerabilities from 2016 to 2022. According to the trend, DL models were first introduced in 2016 for vulnerability detection, since then the use of RNNs for vulnerability detection showed an upward trend. 
The graph also demonstrates a rising trend in using GNNs for vulnerability detection from 2020 to 2022. This can be because GNNs are more powerful than RNNs in detecting vulnerabilities, as they can capture more meaningful and semantic representations of input source code. Since vulnerability types often have complex structures, GNNs are an excellent fit for detecting hidden structural information.


Table \ref{tbl:DL_dist} shows the distribution of DL models used in primary studies. As shown in the table, LSTM is the most frequently used recurrent model, appearing in 8 studies. BiLSTM and BGRU are also popular models with 8 and 6 studies respectively. It is also observable that GGNN is the most prevalent graph-based model, appearing in 4 studies. GCN, GAT, and DR-GCN are also commonly used accounting for 4, 3, and 2 studies respectively. The presence of these models highlights the importance of capturing graph structures and relationships between code elements in vulnerability detection. Attention models were also used in 7 primary studies. Attention mechanisms allow models to pay more attention to specific parts of the code or input that are more likely to contain vulnerability-related patterns. This ability to localize relevant information helps identify and understand the factors contributing to vulnerabilities more effectively. CNNs are used in 6 studies. While not as prevalent as recurrent or graph models, CNNs are still considered effective for capturing local patterns and features in vulnerability detection tasks. General models like DBN, Auto Encoders, Memory Neural Networks, GAN, and Para2Vec are also used to a lesser extent, indicating the exploration of diverse deep learning techniques in vulnerability detection. Transformers are the least frequent family of DL models used in primary studies. This is because they have been recently introduced for software vulnerability detection. Transformers are effective in this domain since they generate contextualized representations for each token in the input sequence. By considering the surrounding tokens and their interactions, transformers capture rich contextual information, which is crucial for understanding vulnerabilities that depend on the overall context of the code or vulnerability-related text.

Table \ref{tbl:ML_dist} shows the distribution of ML models used in primary studies. As shown in the figure, Random Forest is the most frequently used ML model, appearing in 7 studies. SVM, Naive Bayes, and Logistic Regression are popular choices, with 6, 5, and 4 occurrences, respectively. N-Gram models are used in 1 study, indicating their application in capturing sequential patterns and language-based features in vulnerability detection. N-Gram models are commonly used for text analysis and have been adapted for code analysis tasks. Distance measures are utilized in 2 studies for vulnerability detection. These metrics help quantify the similarity or dissimilarity between code elements or features, enabling the identification of potentially vulnerable code segments based on their proximity to known vulnerabilities.

\input{Tables/table8}
\input{Tables/table9}

\mybox{Answer to RQ3}{red!40}{gray!10}{
(1) 79.1\% of primary studies use DL models for vulnerability detection while merely 16.4\% of the primary studies use classic ML models.\\

(2) RNNs and GNNs are by far the most popular DL-based models in software vulnerability detection accounting for 28\% and 22\% of primary studies.\\

(3) LSTM is the most popular architecture in RNN-based models.

(4) Graph-based models are the second most popular models used in software security vulnerability detection accounting for 15 studies. In this family, GGNN is the most popular architecture.

(5) Besides DL models, ML models are popular for software vulnerability detection. Random Forest is the most popular model accounting for 7 studies. 
}

%% file: Tables/table8.tex
\begin{table}[h]
\caption{Distribution of DL models in primary studies.}
\setlength{\tabcolsep}{3pt}
\scalebox{0.9}{
\begin{tabular}{lllll}
\toprule
Category                          & Model Name            & \# Studies & Total    & References           \\
\midrule
\multirow{1}{*}{Recurrent Models} & LSTM                   & 8         & \multirow{1}{*}{31(21)}    &    \cite{dam2018automatic,
dam2017automatic,
wang2016automatically,
tian2020bvdetector,
liu2020cd,
liu2019deepbalance,
ziems2021security,
zhuang2022just}         \\
                                  & BiLSTM                  & 8       &                       &   \cite{jeon2021autovas,
tian2020bvdetector,
3429444,
zou2019mu,
lin2017poster,
lin2019software,
li2021vuldeelocator,
li2018vuldeepecker}         \\
                                  & BGRU                  & 6       &                        &      \cite{jeon2021autovas,
tian2020bvdetector,
yan2021han,
3429444,
li2021sysevr,
li2021vuldeelocator}      \\
                                  & GRU                   & 4       &                        &  \cite{jeon2021autovas,
tian2020bvdetector,
3360588,
hin2022linevd}         \\
                                  & RNN                   & 3       &                        &      \cite{tian2020bvdetector,
le2018maximal,
filus2020random}      \\
                                  & BRNN                   & 2       &                        &      \cite{tian2020bvdetector,
le2018maximal}      \\
                                  \hline
\multirow{1}{*}{Graph Models}     &  GGNN                  & 4       &        \multirow{1}{*}{18(15)}                   &     \cite{wang2020combining,
dinella2020hoppity,
ding2022velvet,
zou2022mvulpreter}    \\
                                & GCN                     & 4         & &     \cite{cheng2021deepwukong,
zeng2021gcn2defect,
ghaffarian2021neural,
li2021vulnerability}        \\
                                  & GAT                   & 3       &                      &    \cite{cheng2021deepwukong,
fu2022linevul,
ghaffarian2021neural}     \\
                                  & DR-GCN                & 2       &                      &      \cite{liu2021combining,
zhuang2020smart}   \\
                                  & RGCN                  & 1       &                      &        \cite{zheng2021vulspg} \\
                                  & FS-GNN                & 1       &                     &      \cite{cao2022mvd}    \\
                                  & K-GNN                 & 1       &                      &       \cite{cheng2021deepwukong}  \\
                                  & DGCNN                 & 1       &                      &      \cite{phan2017convolutional}   \\
                                  & GGRN                  & 1       &                      &      \cite{zhou2019devign}   \\
                                  \hline
Attention Models  & -    & 7         & \multirow{1}{*}{7}   &           \cite{le2021deepcva,
3360588,
zou2019mu,
duan2019vulsniper,
cheng2022path,
liu2021smart,
zhang2022reentrancy} \\
\hline
Convolutional Models              & CNN                   & 6       & 6                     &      \cite{hoang2019deepjit,
yan2021han,
3429444,
8594942,
li2017software,
filus2020random}  \\
\hline
\multirow{1}{*}{General Models}   &   DBN                   & 1       &         5(4)             &     \cite{wang2016automatically}     \\
                                  & Auto Encoders         & 1       &                     &       \cite{le2018maximal}   \\
                                  & Memory Neural Network & 1       &                      &       \cite{hoang2019deepjit}  \\
                                  & GAN                   & 1       &                      &       \cite{3327890} \\
                                  & Para2Vec              & 1       &                      &       \cite{le2018maximal} \\
                                  \hline
Transformers                      &     Seq2Seq Transformer    & 1       &     4(4)                &     \cite{chen2021neural}      \\
                                  &     Graph CodeBERT    & 1       &                     &     \cite{wu2021peculiar}      \\
                                  &     CodeBERT    & 1       &                     &     \cite{ni2022best}      \\
                                  &     HGT    & 1       &                     &     \cite{yang2022source}      \\
\hline
SUM              & -                   & -       & 71(49)    &     -  \\
\bottomrule
\end{tabular}
}
\label{tbl:DL_dist}
\end{table}

%% file: Tables/table9.tex
\begin{table}[H]
\caption{Distribution of ML and other models in primary studies.}
\begin{tabular}{lllll}
\toprule
Category         & Model Name               & Studies & Total & References\\
\midrule
\multirow{1}{*}{Classic ML Models} & Random Forest & 7 & \multirow{1}{*}{35(11)} & \cite{chen2020machine,
sabetta2018practical,
zhou2017automated,
kronjee2018discovering,
pascarella2019fine,
scandariato2014predicting,
Vuldiggerpaper}\\
                 & SVM                      & 6       &       & \cite{chen2020machine,
sabetta2018practical,
zhou2017automated,
scandariato2014predicting,
perl2015vccfinder,
riom2021revisiting} \\
                 & Naive Bayes              & 5      &       & \cite{chen2020machine,
zhou2017automated,
kronjee2018discovering,
pascarella2019fine,
scandariato2014predicting} \\
                 & Logistic Regression      & 4       &       & \cite{sabetta2018practical,
zhou2017automated,
kronjee2018discovering,
pascarella2019fine} \\
                 & K-NN                  & 3       &       & \cite{chen2020machine,
zhou2017automated,
scandariato2014predicting} \\
                 & Gradient Boosting        & 2       &       & \cite{chen2020machine,
zhou2017automated}  \\
                 & Decision Tree            & 2       &       & \cite{kronjee2018discovering,
scandariato2014predicting} \\
                 & AdaBoost                      & 2       &       & \cite{chen2020machine,
zhou2017automated}\\
                 & PCA                      & 1       &       & \cite{yamaguchi2011vulnerability} \\
                 & Kernel Machine           & 1       &       & \cite{nguyen2020deep} \\
                 & ADTree                   & 1       &       & \cite{pascarella2019fine} \\
                 & MLP                      & 1       &       & \cite{pascarella2019fine}\\
                 \hline
Language Models  & N-Gram                   & 1       & 1     & \cite{shippey2019automatically}\\
\hline
Distance Metrics & Distance Measure         & 2       & 2   & \cite{yamaguchi2013chucky, huang2021hunting}\\
\hline
SUM & -         & -       & 38(14)   & -\\
\bottomrule
\end{tabular}
\label{tbl:ML_dist}
\end{table}

%% file: Sections/RQ4.tex
\subsection{RQ4. What is the most frequent type of vulnerability covered in primary studies?}

Software vulnerability detection datasets support different vulnerability types. For example, NVD and SARD benchmark together support 96 types of vulnerabilities. This research question intends to summarize what is the most popular vulnerability types covered by primary studies and what is their frequency. Table \ref{vulnerabilityTypes} shows the statistics regarding the vulnerability types. The column CWE-Type indicates the type of CWE\footnote{\url{https://cwe.mitre.org/}}. There also exists a numerical score for some types of CWEs, which indicate the weakness’ severity. A larger value indicates a higher level of dangerousness and severity\footnote{\url{https://cwe.mitre.org/top25/archive/2021/2021_cwe_top25.html}}. Please note that some frequent types do not have a CWE score, so we denote them as ``-''. There are many categories on the CWE website for vulnerability categorization including \textit{categorization by software development}, \textit{categorization by Hardware design}, and \textit{categorization by research concepts}. The categorization shown in Table \ref{vulnerabilityTypes} is based on \textit{categorization by research concepts} as this categorization is a perfect match for vulnerability types reported in primary studies. 

Table \ref{vulnerabilityTypes} indicates that the vulnerability category that receives the highest attendance is related to the improper control of a resource through its lifetime (CWE-664), with a total of 42 studies (18 unique studies). This category primarily involves managing a system's resources, which are created, utilized, and disposed of according to a predefined set of instructions. When a software system fails to follow these guidelines for resource usage, it can lead to unexpected behaviors that create potentially hazardous situations. Attackers can take advantage of these situations to exploit the software system for their own purposes. It is observable that CWE-119 \cite{russell2018automated, liu2020cd, nguyen2020deep, cheng2021deepwukong, yan2021han, fu2022linevul, le2018maximal, cao2022mvd, chen2021neural, lin2019software, filus2020random, li2018vuldeepecker, duan2019vulsniper, wang2020combining} is the most frequent vulnerability type addressed by the primary studies. This vulnerability occurs when a software system attempts to access or write to a memory location outside the permitted boundary of the system's buffer. Attackers can exploit this vulnerability by controlling memory locations and executing their own code or commands, effectively manipulating the system's memory. Although the score for this vulnerability type is not high, the frequency of primary studies addressing it can be a valuable indicator of its significance in terms of detecting and addressing the vulnerability. Within this category, the most severe vulnerability type with a score of 65.93 is CWE-787, which is discussed in 5 primary studies. This vulnerability is considered severe and critical because it can result in the corruption of data, system crashes, or the execution of malicious code. It occurs when a software system attempts to write data beyond the intended buffer, either before the beginning or past the end of the buffer. CWE-22 is another frequent vulnerability type addressed by primary studies accounting for 4 primary studies \cite{fu2022linevul, ghaffarian2021neural, cheng2021deepwukong, cao2022mvd}. This vulnerability is referred to as ``Path Traversal'', where attackers exploit special elements, such as ``..'' or ``/'', to construct their own path and gain unauthorized access to restricted locations. This vulnerability is particularly critical because attackers can use it to modify restricted files or directories in vulnerable software systems, potentially leading to system crashes by uploading malicious code or commands. In critical financial software systems, attackers can even gain access to customers' bank account information. Given the severity of this vulnerability type with a score of 14.69 and the frequency of papers covering it, detecting and addressing this vulnerability is of utmost importance. Therefore, more advanced ML/DL models are needed to effectively detect this type of vulnerability. The least frequent vulnerability type of this family is CWE-120 \cite{russell2018automated, cao2022mvd} which is a classic buffer overflow, accounting for 2 primary studies. This vulnerability occurs when the software attempts to copy a value to the output buffer without first validating its size. If the range of the value is too large for the length of the output buffer, a buffer overflow can occur. While the detection of buffer overflow can be challenging in some cases, static bug detection tools have already addressed this vulnerability, and detection methods are currently available. 

\input{Tables/table10}

\textit{Improper Neutralization - (CWE-707)} 
Is the second major family of vulnerability types covered by 16 primary studies, including 9 unique primary studies. In this type, the attackers exploit input and output data when they are malformed or not validated properly. There are several scenarios that can lead to data neutralization weaknesses. The first scenario involves checking whether input and output data are safe, while the second scenario involves filtering the input and output data to ensure that any data transformation is done safely. The third scenario involves preventing external attackers from directly manipulating the input and output data, while the fourth scenario relates to the lack of processing input and output data in any circumstances. All of these scenarios can be the root causes of data neutralization weaknesses. As can be seen, CWE-20 is the most frequent type of vulnerability accounting for 6 primary studies. CWE-20 refers to a situation where input validation is not done properly in software systems, making them vulnerable to attacks by malicious individuals who can exploit input data. This occurs when the input data is not verified to be safe or in line with the predefined specifications. The severity and occurrence of this issue are significant, highlighting the need for its detection as it can pose critical risks. CWE-78 is the second major vulnerability type covered by 5 primary studies~\cite{wang2020combining, cheng2021deepwukong, ghaffarian2021neural, li2021vuldeelocator, yang2022source}. This category of security vulnerability pertains to OS command injection, in which an external attacker can construct an OS command by using input data from components that have not been adequately verified. The attacker can then execute harmful commands, potentially causing the system to behave unexpectedly or crash, and putting it in a hazardous state.

\mybox{Answer to RQ4}{red!40}{gray!10}{
(1) The most frequent type of vulnerabilities covered in primary studies is \textit{Improper Control of a Resource Through its Lifetime-(CWE-664)} accounting for 18 unique primary studies. \textit{CWE-119: Improper Restriction of Operations within the Bounds of a Memory Buffer} is the most frequent type of vulnerability in this category accounting for 14 primary studies, and \textit{CWE-787: Out-of-bounds Write} comes subsequently covered by 5 primary studies. The least frequent type of vulnerability is CWE-120 covered by 2 studies.\\
(2) \textit{Improper Neutralization(CWE-707)} is the second major family of vulnerability types covered by 9 unique primary studies in total. In this family, \textit{CWE-20: Improper Input Validation} is the most frequent type covered by 6 primary studies, and \textit{CWE-78: Improper Neutralization of Special Elements used in an OS Command (`OS Command Injection')} comes next with 5 primary studies.\\
(3) Some vulnerability types have a high CVSS score while not sufficiently addressed by existing vulnerability detection studies including but not limited to CWE-79 and CWE-89.
}

%% file: Tables/table10.tex
\begin{table}[t]
\caption{Different vulnerability types covered in primary studies.}
\scalebox{0.8}{
\begin{tabular}{p{0.1\linewidth}cccp{0.07\linewidth}l}
\toprule
Category & CWE-Type & Severity Score  & \#Studies & Total & References\\
\midrule
\multirow{1}{*}{CWE-664} & CWE-119 & 5.84 & 14 & \multirow{1}{*}{42(18)} & \cite{russell2018automated, liu2020cd, nguyen2020deep, cheng2021deepwukong, yan2021han, fu2022linevul, le2018maximal, cao2022mvd, chen2021neural, lin2019software, filus2020random, li2018vuldeepecker, duan2019vulsniper, wang2020combining} \\
         & CWE-787  & 65.93 & 5         &         &  \cite{wang2020combining, cheng2021deepwukong, fu2022linevul, cao2022mvd, yang2022source} \\
         & CWE-22  & 14.69 &4         &             &  \cite{wang2020combining, cheng2021deepwukong, fu2022linevul, ghaffarian2021neural}  \\
         & CWE-125  & 24.9 & 4         &            &  \cite{wang2020combining, cheng2021deepwukong, cao2022mvd, chen2021neural} \\

         & CWE-400  & - & 4         &             &  \cite{wang2020combining, zou2019mu, cheng2021deepwukong, yang2022source} \\
         & CWE-200  & 4.74 & 3         &            &  \cite{wang2020combining,fu2022linevul, chen2021neural} \\
         & CWE-121  & - & 3         &             &  \cite{russell2018automated, yan2021han, cao2022mvd} \\
         & CWE-122  & - & 3         &             &  \cite{3172102, cao2022mvd, russell2018automated} \\
         & CWE-120  & - & 2         &             &  \cite{russell2018automated, cao2022mvd} \\
         \hline
\multirow{1}{*}{CWE-707}   & CWE-20  & 20.47 & 6 & \multirow{1}{*}{16(9)}               &     \cite{russell2018automated, wang2020combining, cheng2021deepwukong, fu2022linevul, chen2021neural, yang2022source}       \\
                                                 & CWE-78  & 19.55 & 5 &           &    \cite{wang2020combining, cheng2021deepwukong, ghaffarian2021neural, li2021vuldeelocator, yang2022source}      \\
                                                 & CWE-89  & 19.54 & 3 &           &       \cite{ghaffarian2021neural, kronjee2018discovering, wang2020combining}   \\
                                                 & CWE-79  & 46.84 & 2 &           &       \cite{kronjee2018discovering, wang2020combining}   \\
\hline                       
\multirow{1}{*}{CWE-682}    & CWE-190     & 7.12 & 5 & \multirow{1}{*}{5} & \cite{wang2020combining, cheng2021deepwukong, fu2022linevul, zou2019mu, yang2022source} \\
\hline
\multirow{1}{*}{CWE-703} & CWE-476  & 6.54 & 4 & \multirow{1}{*}{4} & \cite{russell2018automated, wang2020combining, cao2022mvd, chen2021neural} \\
    \hline
\multirow{1}{*}{CWE-284} & CWE-284 & - & 2 & \multirow{1}{*}{2} & \cite{aivatoglou2021tree, chen2021neural} \\
    \hline
\multirow{1}{*}{CWE-691} & CWE-362  & - & 2 & \multirow{1}{*}{2} & \cite{chen2021neural, zou2019mu} \\
\hline
\multirow{1}{*}{CWE-1215} & CWE-129 & - & 1 & \multirow{1}{*}{1} & \cite{yang2022source} \\
\hline
\multirow{1}{*}{-} & CWE-789 & - & 1 & \multirow{1}{*}{1} & \cite{yang2022source} \\
\hline
\multirow{1}{*}{SUM} & - & - & - & 73(21) & - \\
\bottomrule
\end{tabular}
}
\label{vulnerabilityTypes}
\end{table}


%% file: Sections/RQ6.tex
\subsection{RQ5. What are possible challenges and open directions in software vulnerability detection?} 
%
We have summarized the challenges from previous studies into five different categories, 
which are discussed as follows. 



\noindent \textbf{Challenge 1: Semantic Representation.} 
The biggest challenge in vulnerability detection through learning is the inadequate modeling of the comprehensive semantics of complex vulnerabilities by current models \cite{dam2018automatic, dam2017automatic, shippey2019automatically, wang2016automatically, jeon2021autovas, phan2017convolutional, hoang2019deepjit, cheng2021deepwukong, zhou2019devign, wu2021peculiar,li2017software, li2021sysevr, li2021vuldeelocator, duan2019vulsniper, zheng2021vulspg, zhuang2022just, cheng2022path, liu2021smart,yang2022source, nguyen2022mando, nguyen2022mando2}. These vulnerabilities often exhibit intricate characteristics and patterns that are not fully captured by existing ML/DL models that treat source code snippets as a linear sequence like natural language, or only partially represent source code snippets. Unlike natural language, the source code of real-world projects contains structural and logical information that must be considered by ML/DL models using AST, data flow, and control flow. Therefore, current ML/DL approaches fall short of identifying complex vulnerability patterns.

\noindent \textbf{Challenge 2: Prediction Granularity.}
The ability of DL models to identify the location of vulnerabilities is influenced by the level of granularity in their inputs. Current DL models use a coarser level of granularity, such as method and file, for vulnerability detection. To achieve finer-grained inputs, program slicing is necessary, but it poses a challenge. The crucial question is how to perform program slicing effectively to eliminate unwanted noise in input data and provide more specific inputs. Current tools \cite{pascarella2019fine,
dinella2020hoppity, fu2022linevul, zou2019mu, cao2022mvd, lin2017poster, li2021vulnerability, ding2022velvet,
zou2022mvulpreter} concentrate on library/API function calls, arithmetic operations, and pointer usages, but this approach is not sufficient since not all vulnerabilities originate from these slicing criteria.

\noindent \textbf{Challenge 3: False Positive Removal.} 
The most commonly used tools for detecting software vulnerabilities and bugs are static analyzers \cite{zhou2017automated,
kronjee2018discovering,
zeng2021gcn2defect,
3360588,
3327890,
ziems2021security,
zou2019mu}. These tools utilize hard-coded rules that are defined by experts to model the runtime behavior of a program without the need for compilation. This approach has several benefits, such as effectively identifying the location of vulnerabilities in source codes, which is challenging in large-scale projects with thousands of files and artifacts. Additionally, static analyzers are used at the early stage of the software development process, which helps reduce software maintenance costs. However, relying on expert-defined rules comes with a high false-positive rate, as these rules may not be generalizable to new vulnerabilities that have intricate and sophisticated program semantics. Another significant issue is that defining and updating rules is a labor-intensive and time-consuming process that requires experts to have in-depth knowledge of emerging vulnerabilities. That is why data-driven vulnerability detection has emerged to overcome the aforementioned challenges \cite{zhou2017automated,
kronjee2018discovering, zeng2021gcn2defect, 3360588, 3327890, ziems2021security, zou2019mu}. 

\noindent \textbf{Challenge 4: Lack of Training Data.}
A significant weakness of DL models, particularly in software vulnerability detection, is their insatiable need for data \cite{chen2020machine,liu2020cd, lin2019software, Vuldiggerpaper, ni2022best}. In domains such as image classification, there is an ample supply of labeled data, making it possible to train DL models effectively. Furthermore, there are many pre-trained models available that can be fine-tuned for detection tasks. However, in software vulnerability detection, data scarcity is a major problem since labeling ground truth information is a challenging task. To obtain training data, multiple online platforms such as Stack Overflow, GitHub, and issue tracking or bug tracking systems are used. While these platforms contain billions of records, the labeling process is difficult and is often done manually. One possible solution is the automatic labeling of data, but this approach is challenging as it often generates many false positives. Additionally, some researchers use unsupervised classification for vulnerability detection, but this method also suffers from limited precision.



\noindent \textbf{Challenge 5: Lack of Model Interpretability.}
Interpretability in DL models refers to the ability to understand and explain the decisions made by the model~\cite{dwivedi2023explainable, minh2022explainable}. In the context of software vulnerability detection~\cite{3429444, li2021vulnerability,liu2021smart}, there are several challenges that make interpretability challenging for DL models. First, source code can be highly complex, especially in large software projects. It often consists of multiple files, functions, and dependencies, making it difficult to extract meaningful and concise explanations from the code~\cite{jiang2022bugs}. Software systems are dynamic and undergo changes over time. Code is often maintained, updated, and refactored, which can introduce complexities in interpreting the decisions made by AI models. The model's explanations may not be applicable to the current version of the code if it has evolved since the model's training.


\mybox{Answer to RQ5}{red!40}{gray!10}{
(1) Current models inadequately capture the comprehensive semantics of complex vulnerabilities, as they fail to consider the structural and logical information present in source code snippets. Existing ML/DL models treat source code as a linear sequence, which limits their ability to identify intricate vulnerability patterns.

(2) DL models often use a coarse level of granularity for vulnerability detection, such as method and file level. Achieving finer-grained inputs requires effective program-slicing techniques to eliminate noise and provide more specific inputs. Current approaches focus on certain slicing criteria, but vulnerabilities can originate from other sources as well.

(3) DL models require a significant amount of labeled data for effective training. However, in software vulnerability detection, there is a scarcity of labeled data due to the challenging task of manual labeling. Automatic labeling approaches often generate false positives, and unsupervised classification suffers from limited precision.
}

%% file: Sections/Limitations.tex
\section{Threats to validity}
\noindent \textbf{External Validity}.
One of the major threats to the external validity of our work is the data collection internal. We collect data in an 11 years old period from 2011 to 2022 to coverall all possible studies published during this period. Another source of threat to external validity is the coverage of the input data types in software vulnerability detection. To tackle this threat, we focused on source code snippets as well as repository data, i.e., data that can be extracted from open source including GitHub and CVE. 

\noindent \textbf{Internal Validity}. 
One of the major threats to the internal validity of our work is the automatic collection of data for analysis. Our technique for data collection is automated at its initial steps in which we may miss some important vulnerability detection papers. In fact, we developed a set of scripts that allowed us to extract papers given an 11-year period. Even though the subsequent steps are manually supervised, still the automatic filtering suffers from this issue. The second important issue with our data collection is possible bias. The root cause of bias is unavoidable disagreements during paper classification as two researchers have worked together to categorize papers based on title, abstract, and content of papers. In case the researchers could not come up with an agreement, the third researcher join to resolve the differences which somehow relaxes the issue. 

\noindent \textbf{Construct Validity}. The major threat to the construct validity of our survey is the level of granularity of the analysis we conducted for each primary study. For each study, we deeply analyzed the artifacts explained in the paper and show their distribution as tables and graphs. For example, in terms of sources for benchmark data, we deeply analyzed 17 sources accounting for 53 primary studies overall. The second threat to the construct validity is the degree of coverage for each primary study. We analyzed each primary study from 5 aspects including input data, input representation, embedding techniques, models, vulnerability types, and whether the study supports the interpretability of vulnerability detection models or not.  


%% file: Sections/conclusion.tex
\section{Conclusion}
In this study, we conducted a systematic survey on 67 primary studies using ML/DL models for software security vulnerability detection. We collected the papers from different journals and conference venues including 25 conferences and 12 journals. Our review is established based on five major research questions and a set of sub-research questions. We devised the research questions in a comprehensive manner where to cover various dimensions of software vulnerability detection. Our analysis of primary studies indicated that there is a booming trend in the growth of using ML/DL models for software vulnerability detection. Our deep analysis of data sources of primary studies revealed that 65.7\% of studies use benchmark data for software vulnerability detection. We also find 6 broad categories of DL models along with 14 classic ML models used in software vulnerability detection. The categories of DL models are classified as recurrent models, graph models, attention models, convolutional models, general models, and transformer models. RNNs are by far the most popular DNNs in software vulnerability detection. Our analysis also finds that RNNs with LSTM cells are the most popular network architectures in recurrent models, accounting for 8 primary studies. In the category of graph models, GGNN is the most popular DL model used by 4 primary studies. Our results on vulnerability types reveal that the most frequent type of vulnerability covered in existing studies is Improper Control of a Resource Through its Lifetime - (CWE-664) accounting for 18 primary studies. In conclusion, we have identified a collection of ongoing challenges that necessitate further exploration in future studies involving the utilization of ML/DL models for software vulnerability detection.